\documentstyle[here,epsfig]{mn2e} 
\def\simlt{\mathrel{\rlap{\lower 3pt\hbox{$\sim$}}\raise 2.0pt\hbox{$<$}}}
\def\simgt{\mathrel{\rlap{\lower 3pt\hbox{$\sim$}} \raise 2.0pt\hbox{$>$}}}

\def\Msun{M_{\odot}}

\def\gtsima{$\; \buildrel > \over \sim \;$}\def\gtsima{$\; \buildrel > \over
  \sim \;$}

\def\ltsima{$\; \buildrel < \over \sim \;$}
\def\gtrsim{\lower.5ex\hbox{\gtsima}}
\def\lesssim{\lower.5ex\hbox{\ltsima}}

\newcommand{\q}{\begin{equation}}
\newcommand{\qa}{\begin{eqnarray}}
\newcommand{\qs}{\begin{eqnarray*}}
\newcommand{\nq}{\end{equation}}
\newcommand{\nqa}{\end{eqnarray}}
\newcommand{\nqs}{\end{eqnarray*}}

\begin{document}

\title[BSSs in dSphs II] 
{Blue straggler stars in dwarf spheroidal galaxies II: Sculptor and Fornax}

\author[M. Mapelli et al.]{M. Mapelli$^{1}$, E. Ripamonti$^{2}$, G. Battaglia$^{3}$, E. Tolstoy$^{4}$, M. J. Irwin$^{5}$, \newauthor B. Moore$^{1}$, S. Sigurdsson$^{6}$ 
\\
$^1$Institute for Theoretical Physics, University of Z\"urich, Winterthurerstrasse 190, CH--8057 Z\"urich, Switzerland; {\tt
mapelli@physik.unizh.ch}\\ 
$^2$Dipartimento di Fisica e Matematica, Universit\`a dell'Insubria, Via Valleggio 11, I--22100, Como, Italy\\
$^3$European Organization for Astronomical Research in the Southern Hemisphere, K. Schwarzschild-Str. 2, 85748 Garching, Germany\\
$^4$Kapteyn Astronomical Institute, University of Groningen, Postbus 800, 9700 AV Groningen, the Netherlands\\ 
$^{5}$Institute of Astronomy, Madingley Road, Cambridge, CB3 0HA, UK\\
$^6$Department of Astronomy and Astrophysics, The Pennsylvania State University, 525 Davey Lab, University Park, PA~16802, US\\}

\maketitle 
\vspace {7cm}

\begin{abstract}
The existence of blue straggler stars (BSSs) in dwarf spheroidal galaxies (dSphs) is still an open question. In fact, many BSS candidates have been observed in the Local Group dSphs, but it is unclear whether they are real BSSs or young stars. Shedding light on the nature of these BSS candidates is crucial, in order to understand the star formation history of dSphs. 
 In this paper, we consider BSS candidates in Sculptor and Fornax. In Fornax there are strong hints that the BSS population is contaminated by young stars, whereas in  Sculptor there is no clear evidence of recent star formation.
 We derive the radial and luminosity distribution of BSS candidates from wide field imaging data extending beyond the  nominal tidal radius of these galaxies. The observations are compared with the radial distribution of BSSs expected from dynamical simulations. In Sculptor the radial distribution of BSS candidates is consistent with that of red horizontal branch (RHB) stars
and is in agreement with theoretical expectations for BSSs generated via mass transfer in binaries. On the contrary, in Fornax the radial distribution of BSS candidates is more concentrated than that of all the considered stellar populations.
This result supports the hypothesis that most of BSS candidates in Fornax are young stars and is consistent with previous studies.
   
\end{abstract}


\section{Introduction}
Blue straggler stars (BSSs) are located above and blue-ward of the main sequence (MS) turn-off  in a color-magnitude diagram (CMD). They apparently burn hydrogen in their core, although their mass is larger than the turn-off mass. Thus, they must have experienced a chemical mixing, which `rejuvenated' their inner layers. 
The mechanism leading to this chemical mixing is one of the main issues about BSSs. The most common theoretical models suggest that the chemical mixing might be the result of a collision between stars (Sigurdsson, Davies \& Bolte 1994 and references therein)  or of mass-transfer in binaries (McCrea 1964). Thus, understanding the formation mechanisms of BSSs requires both  models of chemical evolution and the study of the  dynamics of stellar systems. 

Several observational and theoretical studies (Sandage 1953; Fusi Pecci
et al. 1992; Ferraro et al. 1993, 1997; Sigurdsson et al. 1994; Zaggia, Piotto \& Capaccioli 1997; Ferraro et al. 2003, 2004;
Sabbi et al. 2004; Hurley et al. 2005; Davies, Piotto \& De Angeli 2004;
Mapelli et al. 2004, 2006, hereafter M04, M06; Lanzoni et al. 2007a,
2007b) support the existence of blue straggler stars (BSSs) in star
clusters, where the tiny spread in the stellar age makes their identification straight forward. 

The possibility that dwarf spheroidal galaxies (dSphs) host BSSs is more controversial. Mateo et al. (1991) and Mateo,
Fischer \& Krzeminski (1995) first indicated the existence of a large
number of stars brighter than the turn-off mass in the Sextans
dSph.
BSS candidates have been found in varying numbers in most dSphs, such as
Sculptor (e.g., Hurley-Keller, Mateo \& Grebel 1999; Monkiewicz et
al. 1999), Draco (Aparicio, Carrera \& Mart\'inez-Delgado 2001) and Ursa Minor (Carrera et al. 2002).
 However, these stars can be either genuine BSSs or ordinary MS stars substantially younger than the bulk of the other stars  (Mateo et al. 1995). 

The existence of real BSSs in dSphs is crucial for understanding the formation and evolution of these galaxies. In fact, BSSs can be confused with young stars ($\lesssim{}2$ Gyr), due to their position in the CMD. If a population of BSSs is mistakenly interpreted as a young MS, we might derive a wrong star formation history for the host galaxy. This risk is particularly strong in dSphs, where the bulk of star formation took place a long time ago ($\sim{}8-10$ Gyr) and the existence of more recent episodes of star formation is still debated. In various dSphs, the observed BSS candidates are often interpreted as young stars (Mateo et al. 1995).

An indication that BSS candidates in dSphs may be real BSSs is given by the work of Momany et al. (2007). The authors recently analysed the BSS candidates of 8 dSphs and found a statistically significant anti-correlation between the relative frequency  of BSS candidates with respect to the horizontal branch (HB) stars 
and the total luminosity of the dSph. Such an anti-correlation is typical of BSSs, as it has already been found in both globular clusters (Piotto et al. 2004) and  open clusters (de Marchi et al. 2006).
If BSS candidates in dwarf galaxies were young MS stars such an anti-correlation would be difficult to explain.

Mapelli et al. (2007, hereafter paper~I) studied by means of both observations and simulations the BSS candidates of Draco and Ursa Minor. These two galaxies are among those dSphs of the Local Group with a predominantly ancient stellar population ($>$ 8-10 Gyr old, see Mateo 1998 - hereafter M98; Hernandez, Gilmore \& Valls-Gabaud 2000; Aparicio et al. 2001; Carrera et al. 2002; Bellazzini et al. 2002). Paper~I showed that the radial distribution and the luminosity distribution of BSS candidates in these systems match the expected properties of `real' BSSs. In particular, in both Draco and Ursa Minor, the radial distribution of BSS candidates is similar to that of red giant branch (RGB) and HB stars. This agrees with theoretical models (McCrea 1964; M04; M06; paper~I), which predict that BSSs in low-density environments, such as dSphs, form mainly via mass transfer in primordial binaries, whose radial distribution is expected to trace the distribution of the other ancient stellar populations. 

In this paper we extend our analysis to the Sculptor and Fornax dSphs. Our aim is to study the observational properties of BSS candidates in these two galaxies and to compare them with theoretical models, in order to  assess whether these stars are genuine BSSs or young stars.

The properties of Sculptor and Fornax are very different from each other and from those of Draco and Ursa Minor.
Sculptor is relatively close to the Milky Way ($\sim{}79$ kpc, M98) and its stellar population is predominantly old ($>10$ Gyr, Kaluzny et al. 1995; Tolstoy et al. 2004; Clementini et al. 2005). Although star formation as recent as 2 Gyr cannot be ruled out (Monkiewicz et al. 1999), there is no unambiguous evidence for intermediate-age stars (2$-$8 Gyr). 

Fornax is relatively distant ($\sim{}138$ kpc; M98), is one of the most luminous and massive companions of the Milky Way, hosts five globular clusters and shows a long and complex star formation history (Stetson, Hesser \&{} Smecker-Hane 1998; Buonanno et al. 1999; Saviane, Held \&{} Bertelli 2000; Pont et al. 2004; Battaglia et al. 2006 - hereafter B06). In particular, the observations suggest the existence of three different stellar populations: i) ancient stars ($> 10$ Gyr), mostly visible as a well populated red HB (RHB), ii) intermediate-age stars ($2-8$ Gyr), including asymptotic giant branch (AGB) and red clump (RC), iii) young stars ($<$1 Gyr), associated with a young blue loop (BL) and possibly with a young MS. In this paper we will refer to a part of the possible young MS as BSS candidates. It is worth  stressing that in Fornax the identification of BSS candidates with young MS stars is commonly accepted, whereas in Sculptor there is no strong evidence for the existence of young stars. The main properties of Sculptor and Fornax are listed in Table~1.

\section{The data}
\begin{figure*}
\center{{
\epsfig{figure=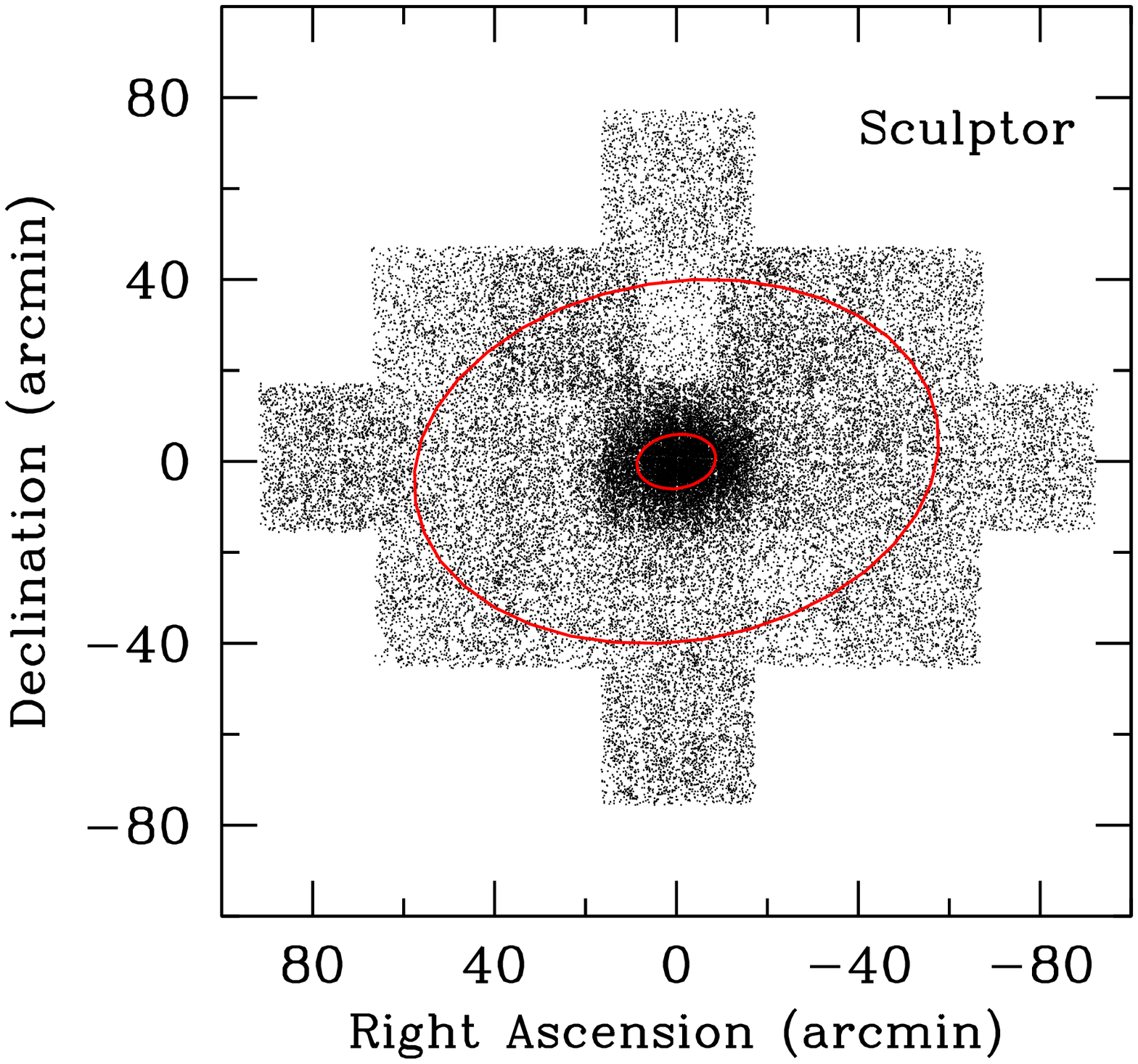,height=8cm}
\epsfig{figure=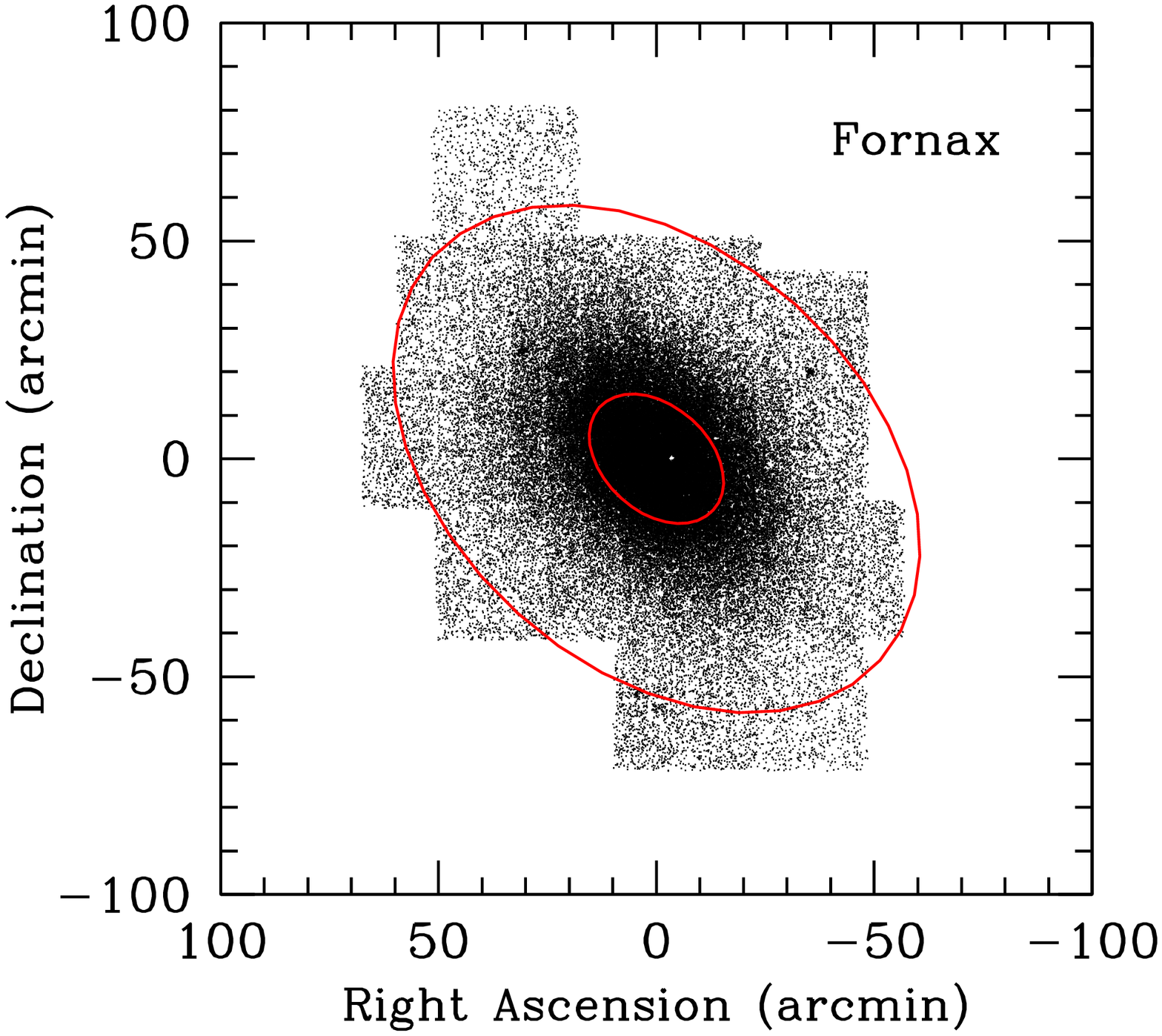,height=8cm}
}}
\caption{\label{fig:fig1} 
Positional map of the stars imaged in Sculptor (left) and Fornax
(right). The concentric ellipses indicate tidal and core radii ($r_t$
and $r_c$; the adopted values are listed in Table~1).
In both cases the origin of the axes coincides with the centre
of the observed galaxy. The North is at the top, and the East on the left-hand side.} 
\end{figure*}
\subsection{WFI data}
The data used here were acquired with the ESO/2.2m Wide Field Imager (WFI) at La Silla between 2003 and 2004 for the Sculptor dSph and in 2005 for the Fornax dSph. A journal of the observations is available in B06 for the Fornax dSph and in Battaglia (2007 - hereafter B07) for Sculptor and 
the coverage for these two galaxies is visible in Fig.~\ref{fig:fig1}. The data reduction was done in a standard way and based on the pipeline processing software 
developed by the Cambridge Astronomical Survey Unit for dealing with imaging 
data from mosaic cameras.  Details of the pipeline processing can be found in 
 Irwin (1985), Irwin \& Lewis (2001) and Irwin et al. (2004).  For more details on the data reduction we refer to B06, Battaglia et al. (2008a) and references therein.

\begin{figure*}
\center{{
\epsfig{figure=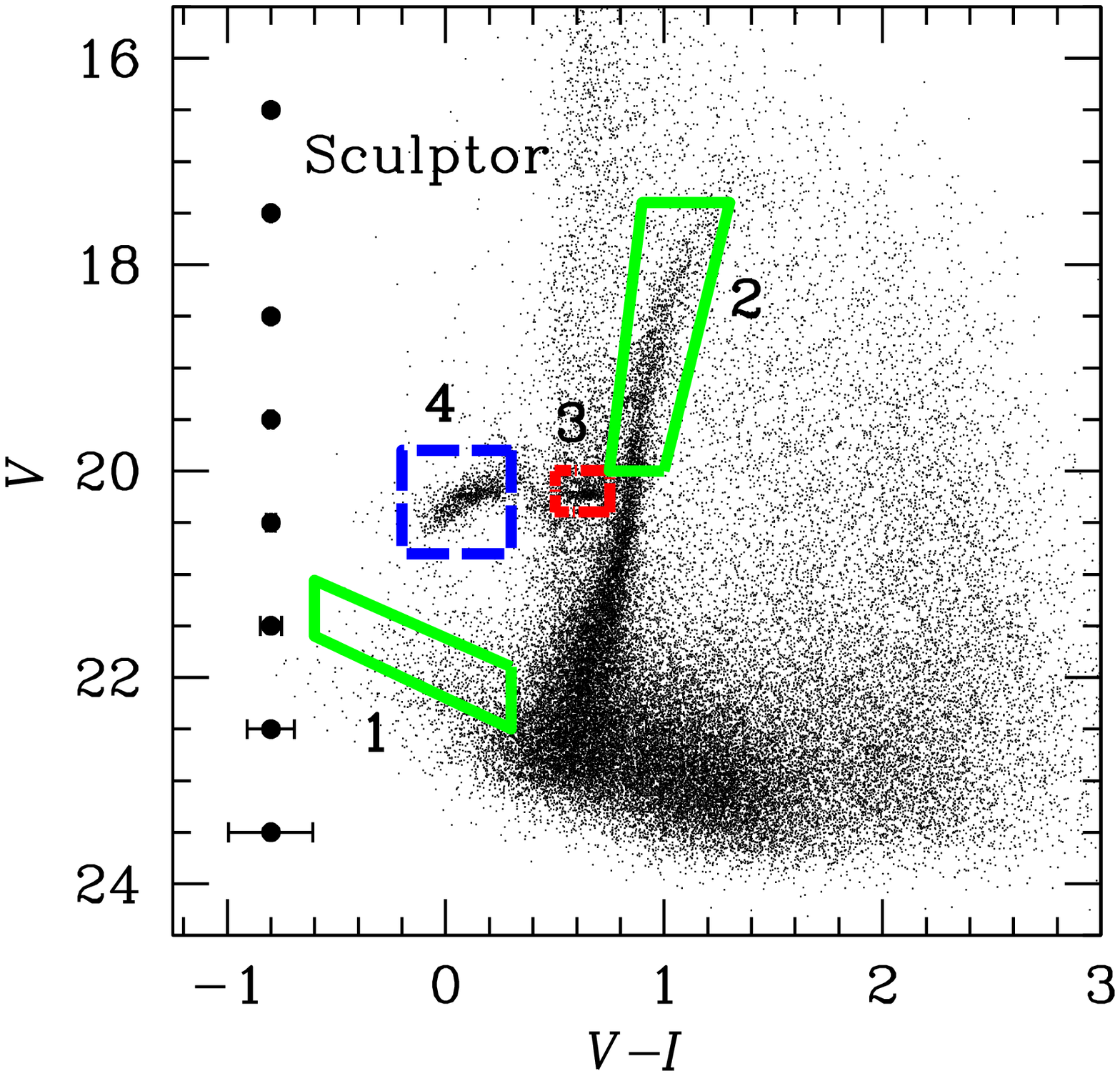,height=8cm}
\epsfig{figure=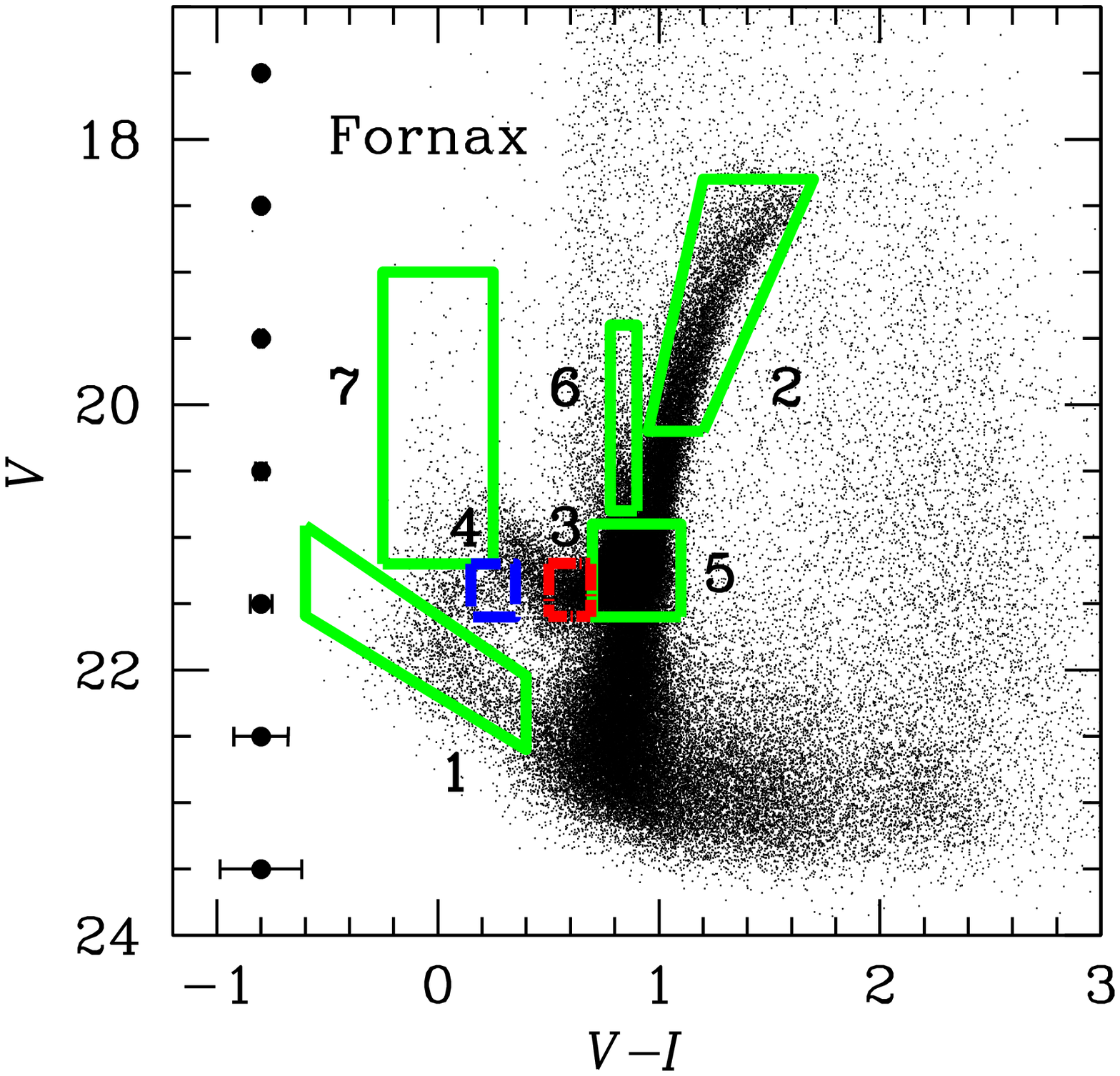,height=8cm}
}}
\caption{\label{fig:fig2} CMD of Sculptor (left-hand panel) and Fornax
(right-hand panel) with stellar population selection boxes overlaid. Boxes indicated 
by the solid line (green on the web) and labeled as 1 and 2 are
the BSSs and RGB stars, respectively. Boxes indicated by the long-dashed dotted line (red on the web) and labeled as 3  are the RHB stars. Boxes indicated by the long dashed line (blue on the web) and labeled as 4 are the BHB stars. In the case of Fornax (right-hand panel), boxes indicated 
by the solid line (green on the web) and labeled as 5, 6  and 7 are
the RC, BL and MS stars, respectively. The error bars refer to the {\it V-I} photometric errors (see B07).}
\end{figure*}

\subsection{Stellar population selection criteria}
From these data 
we selected three different populations: BSS candidates, RGB
and HB stars. The HB stars are $>$10 Gyr old and therefore represent the ancient 
component in these galaxies; the RGB contains stars of any age (but $>$1 Gyr old), and therefore represents the overall stellar population of these galaxies. Among the HB stars, we also distinguish between a red HB (RHB) and a blue HB (BHB)\footnote{The number of observed BHB stars in Fornax is quite low (408), introducing large statistical errors. Furthermore, the BHB sample is difficult to distinguish from young MS stars. However, we decided to consider also the BHB stars in Fornax, for symmetry with Sculptor.} component. 
The regions of the CMD we associate with BSS candidates, RGB, RHB and BHB stars are
indicated in Fig.~\ref{fig:fig2} as boxes 1, 2, 3 and 4, respectively.
 In the case of Fornax we also consider RC, BL and young MS stars, indicated in the right-hand panel of Fig.~\ref{fig:fig2} as boxes 5, 6 and 7, respectively. RC stars are representative of the intermediate-age population (2$-$8 Gyr) in Fornax, whereas young MS\footnote{The number of young MS stars is also quite low (482) and they may partially overlap with BHB stars. However, it is essential to compare them with BSS candidates (see next Section).}  and BL stars are thought to be young stars ($<1$ Gyr). The stars that we define here as `young MS stars' are those stars which surely belong to a young ($<1$ Gyr) MS, because they are above the HB and cannot be confused with real BSSs (although there may be some contamination from BHB  stars for $V\sim{}21.2$). In the literature (B06 and references therein) the young MS stars (box~7 in the right-hand panel of Fig.~\ref{fig:fig2}) and the BSS candidates (box~1 of the right-hand panel of Fig.~\ref{fig:fig2}) are often considered as a single population of young MS stars. Here, we make this distinction because we want to check whether some of the stars in box~1 (or, unlikely, all of them) are not young stars but real BSSs.
For selecting RGB, RHB, BHB, BL and RC stars we adopt the same criteria as in B07 for Sculptor and in B06 for Fornax. The only exception is represented by RGB stars in Sculptor, for which we select a narrower box than in B07, in order to minimize the foreground contamination. Finally, when selecting BSS candidates in both Sculptor and Fornax, we require to be above the 50 per cent completeness limit in $V$ and in $I$. These limits are $V=23.0$, $I= 22.2$ for Sculptor, and  $V=23.7$, $I= 22.2$ for Fornax. Because of this requirement, we miss the faintest BSS candidates. However, most of our results (and in particular the radial distributions) are unaffected by this fact.


\begin{table*}
\begin{center}
\caption{Galaxy properties} \leavevmode
\begin{tabular}[!h]{llllllllll}
\hline
Galaxy
& $d$$^{\rm a}$ (kpc)
& $\alpha{}_{2000}$$^{\rm b}$
& $\delta{}_{2000}$$^{\rm b}$
& $r_{c}$$^{\rm c}$ (arcmin)
& $r_{t}$$^{\rm c}$ (arcmin)
& $\sigma_c$ (km s$^{-1}$)$^{\rm d}$
& $W_0$$^{\rm e}$
& $c$$^{\rm e}$
& ellipticity$^{\rm f}$ \\
\hline
Sculptor       & 79 & $1^h0^m09^s$ & $-33^\circ{}42'30''$& 8.7 & 58.1 & 10.0 & 2.6 &  0.82  & 0.32 \\
Fornax        & 138 & $2^h39^m52^s$ & $-34^\circ{}30'49''$& 17.6 & 69.1 & 10.5 & 1.2 &  0.60  & 0.31 \\
\noalign{\vspace{0.1cm}}
\hline
\end{tabular}
\end{center}
\footnotesize{ $^{\rm a}$ We assume distance moduli of 19.54 (Sculptor) and
  20.70 (Fornax), from  M98. $^{\rm b}$ Right Ascension and Declination of the centre of mass of the galaxy are from B07 and B06 for Sculptor and Fornax, respectively. $^{\rm
c}$Core radius ($r_c$) and tidal radius ($r_t$) are from
B07 and from  B06 for Sculptor and Fornax,
respectively. The values of $r_c$ and $r_t$ adopted for Sculptor (B07) are slightly different from the ones generally adopted (M98 and references therein), but they allow a better match with the simulations discussed in Section 5. The main results are unchanged when we adopt the values of $r_c$ and $r_t$ in  M98. $^{\rm d}$Core velocity dispersion of the dSph adopted in the simulations, consistent with Battaglia et al. (2008b) and with B06 for Sculptor and Fornax, respectively.  $^{\rm e}$Central adimensional potential ($W_0$)
and concentration [$c={\rm log_{10}}(r_t/r_c)$] are derived from our simulations. $c$ is
consistent with B07 for Sculptor and with
B06 for Fornax.  $^{\rm f}$Ellipticities are from  M98.}
\end{table*}
\begin{figure*}
\center{{
\epsfig{figure=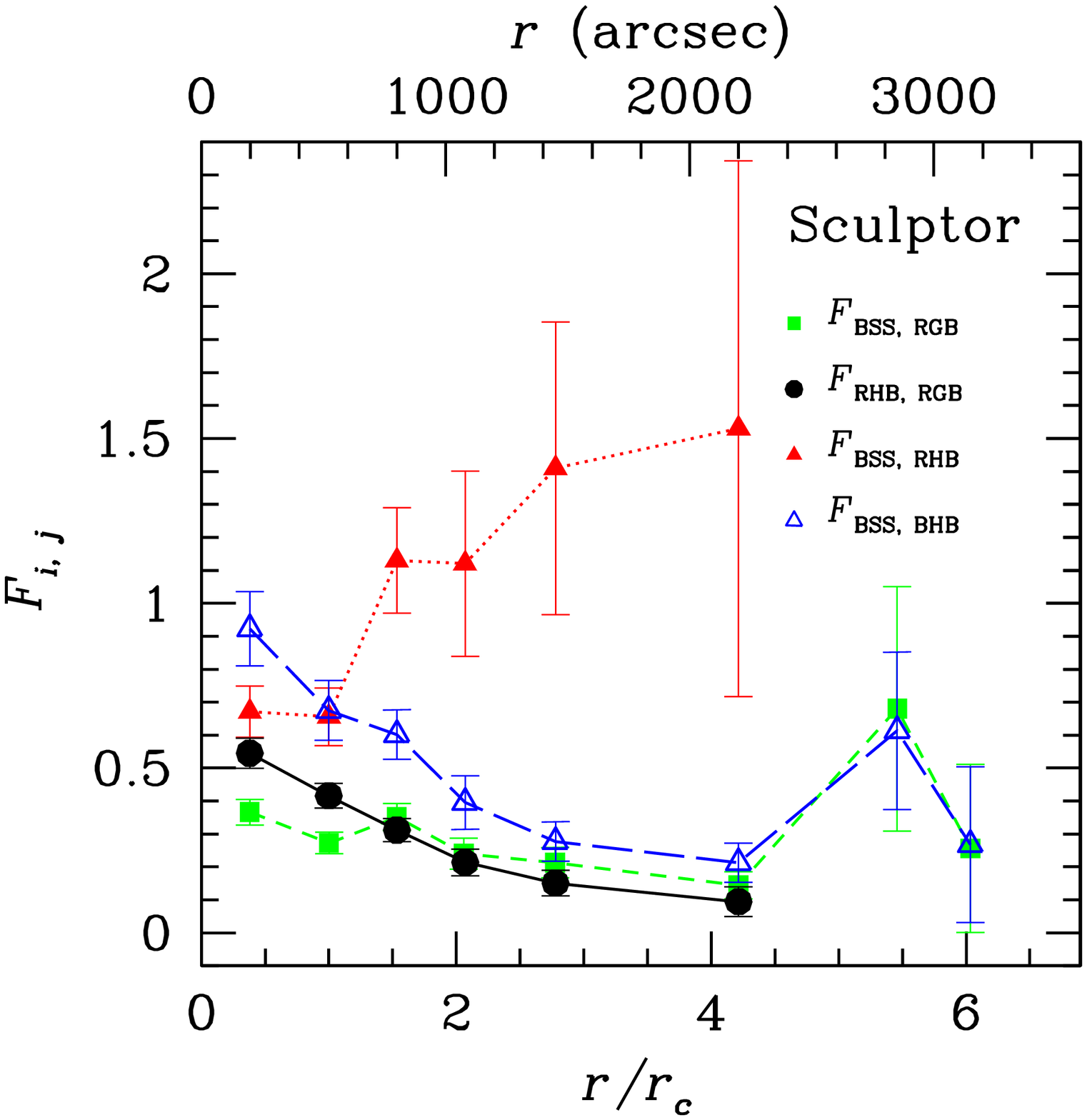,height=8cm}
\epsfig{figure=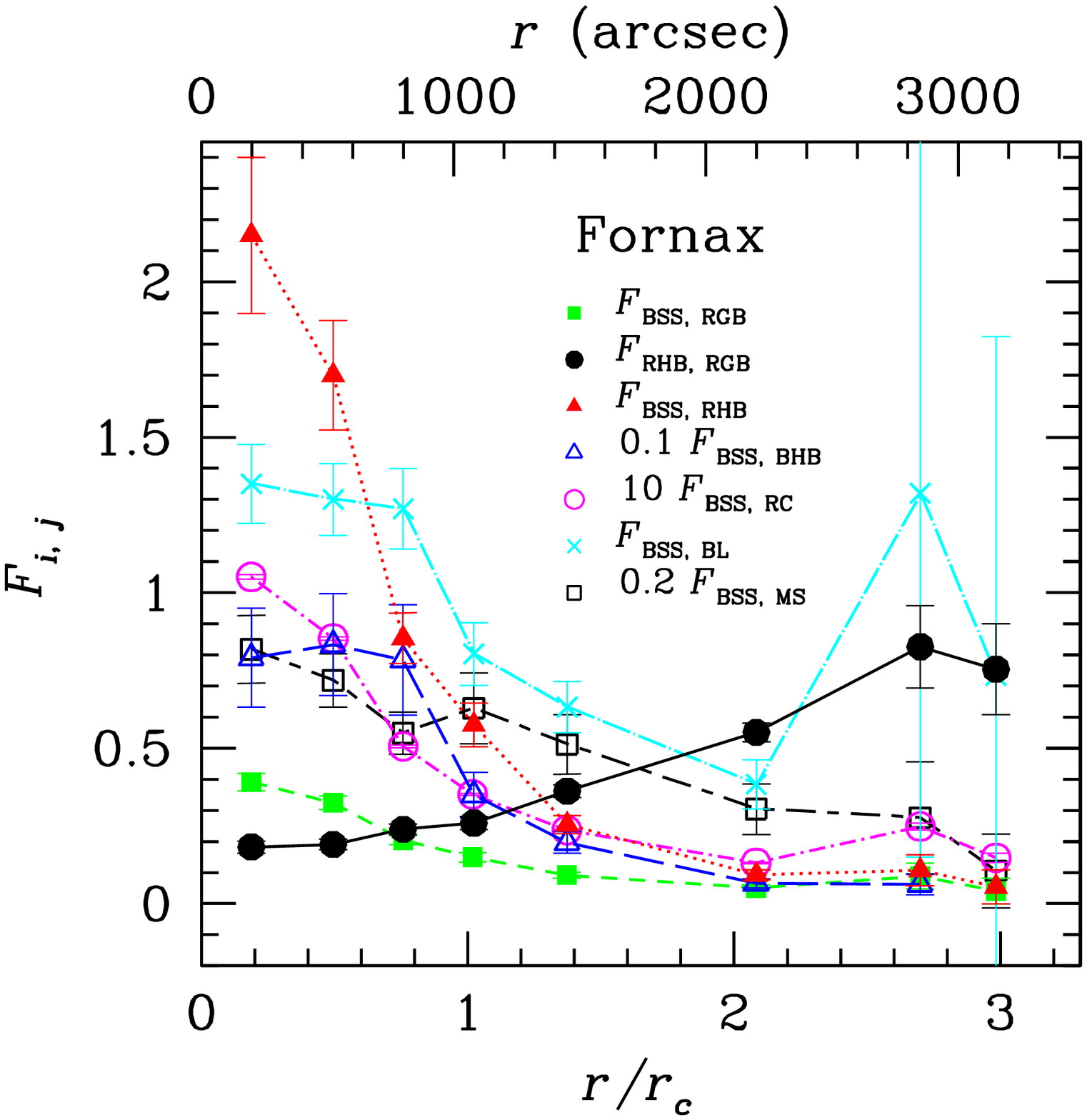,height=8cm}
}}
\caption{\label{fig:fig3}
Observed relative frequency of BSSs normalized to RGB stars ($F_{\rm BSS, RGB}$, filled squares connected by short dashed line, green on the web), to RHB stars ($F_{\rm BSS, RHB}$, filled triangles connected by dotted line, red on the web) and to BHB stars ($F_{\rm BSS, BHB}$, open triangles connected by long dashed line, blue on the web). The observed relative frequency of RHB stars normalized to RGB stars ($F_{\rm RHB, RGB}$) is shown as filled circles connected by solid black line. The left-hand panel refers to Sculptor, the right-hand panel to Fornax. For Fornax we also show the observed relative frequency of BSSs normalized to BL stars ($F_{\rm BSS, BL}$, crosses connected by long-dashed dotted line, cyan on the web), to young MS stars ($F_{\rm BSS, MS}$, open squares connected by long-dashed short-dashed line, black on the web) and to RC stars ($F_{\rm BSS, RC}$, open circles connected by short-dashed dotted line, magenta on the web).
In the case of Fornax  $F_{\rm BSS, BHB}$ ($F_{\rm BSS, MS}$) and the corresponding error bars have been divided by 10 (5), in order to make more readable the Figure. For the same reason, $F_{\rm BSS, RC}$ and the corresponding error bars have been multiplied by 10. All the radial distributions have been corrected for foreground contamination. Error bars account for Poissonian statistics, uncertainties in foreground subtraction and photometric errors (see Section~3).
}
\end{figure*}

\section{Radial distribution of BSS candidates}
The radial distribution of different stellar populations in a galaxy often provides useful insights on the evolution of the system. It has long been known that dSphs 
exhibit radial variations in their stellar population mix, with their blue stars being less spatially concentrated than the red ones (see Harbeck et al. 2001 
for a sample of Local Group dSphs), and this has been interpreted as an age/metallicity gradient. Tolstoy et al. (2004) showed that in Sculptor the BHB and metal-poor stars have a less concentrated 
spatial distribution than the RHB and metal-rich stars. This, combined with the fact that the Sculptor stars are predominantly old ($>$ 10 Gyr old), indicates a metallicity gradient within the ancient stellar population of Sculptor. In Fornax the properties of the 
stellar population mix also change with radius, but over a different range of ages than in Sculptor: ancient ($>$10 Gyr old) and metal-poor stars have a more extended distribution than intermediate-age (2-8 Gyr old) and metal-rich stars. Candidate young MS stars and  BL stars ($<$1 Gyr old) have an even more concentrated distribution. 
 In general, in those dwarf galaxies where there is clear evidence of ongoing star formation or of the presence of young stars, the younger stars appear more centrally concentrated than the older ones (as stated for the first time by Baade \&{} Gaposchkin 1963 in the case of IC~1613; see also Skillman et al. 2003 and references therein). This might indicate that more recent star formation occurs preferentially in the inner regions.


For the BSSs a different radial distribution may indicate a different formation mechanism. Recent dynamical simulations (M04; M06) show that, in globular clusters, a centrally concentrated BSS population can be associated with a collisional origin, whereas BSSs formed by mass transfer in binaries are expected to follow the same radial distribution as the total stellar light. In dwarf galaxies, where the low density makes stellar collisions unlikely, BSSs can form only via mass transfer and thus their radial distribution is expected to be similar to that of  other stellar populations, representative of the light profile.
Paper~I shows that the radial distribution of BSS candidates in Draco and Ursa Minor is similar to (or even less concentrated than) that of RGB and HB stars, supporting the idea that BSS candidates are `real' BSSs.
In this paper we carry out the same kind of analysis in the case of Sculptor and Fornax.

A useful tool to compare the radial distributions of two different populations is the relative frequency (or related quantities), which has already been used in previous studies (see Ferraro et al. 1997 and references therein). The relative frequency $F_{i,j}(r)$ of a given stellar population $i$ with respect to another population $j$ is defined as the  ratio between the number $N_i(r)$ of stars belonging to the population $i$ found in the radial bin with average radius\footnote{All the references to `radii' in this paper mean projected elliptical radii. The elliptical radius of a point $(x,y)$ is
$r_{ell}(x,y)^2 = x^2 + [y/(1-e)]^2$, where $e$ is the ellipticity of
the considered galaxy, and the galaxy is assumed to be centered on the
origin, with its major axis aligned with the x-axis.} $r$ and the number $N_j(r)$ of stars belonging to the population $j$ found in the radial bin with average radius $r$, that is:
\begin{equation}
F_{i,j}(r)=\frac{N_i(r)}{N_j(r)}.
\end{equation}

In Fig.~\ref{fig:fig3} we consider the relative frequency of BSSs versus RGB ($F_{\rm BSS,RGB}$), versus RHB ($F_{\rm BSS,RHB}$) and versus BHB ($F_{\rm BSS,BHB}$) stars. We also consider the relative frequency  of RHB versus RGB stars. In the case of Fornax we also plot the relative frequency of BSSs versus BL ($F_{\rm BSS,BL}$), versus young MS ($F_{\rm BSS,MS}$) and versus RC ($F_{\rm BSS,RC}$) stars. Thus, in the case of Fornax we compare the radial distribution of BSSs with that of populations which are representative of all the different ages found in this dSph, i.e. ancient stars (represented by RHB and BHB), intermediate-age stars (RC) and young stars (BL, MS).
Tables 2 and 3 report the number counts for the considered populations in the case of Sculptor and Fornax, respectively.
Results were corrected for the contamination from foreground (and, to a minor extent, background) objects, using the same method as described in the Appendix A of Paper I; the results are consistent with those obtained by B06 for Fornax and by B07 for  Sculptor.
 Finally, the errors reported in Fig.~\ref{fig:fig3} and in Tables 2 and 3 have been calculated as:
\begin{equation}\label{eq:newerr}
\sigma{}=\left(\sigma{}_{\rm Pois}^2+\sigma{}_{\rm for}^2+\sigma{}_{\rm phot}^2\right)^{1/2},
\end{equation}
where $\sigma{}_{\rm Pois}$, $\sigma{}_{\rm for}$ and $\sigma{}_{\rm phot}$ are the terms due to Poissonian error, foreground contamination and photometric error, respectively. $\sigma{}_{\rm phot}$ has been derived with the bootstrap technique\footnote{In the bootstrap we generate a large number (generally, 1000) of realisations of the observed star catalogue: for each of the observed stars,
the $V$ and $I$ values going into a realisation are drawn randomly from
Gaussian distributions centred on the observed magnitudes, whose width
is given by the relevant photometric error. For each realisation we determine the number count of stars inside the
various color-magnitude selection boxes. Then, we use the  average and variance of these number counts as
estimates of the number of stars within a box, and of the error that
is introduced by photometric uncertainties, respectively.}.

\subsection{Sculptor}
 In Sculptor $F_{\rm BSS,RHB}$ is marginally consistent with a flat distribution: BSSs look slightly less concentrated than RHB stars, but the error bars are quite large. BSSs and RHB stars are slightly more concentrated than RGB stars.
 Furthermore, BSSs are more concentrated than BHB stars. 
This suggests that BSS candidates in Sculptor are mostly connected with the red, metal-rich population of this galaxy.
The similarity between the radial distribution of BSSs and that of RGB and RHB stars also favours the scenario in which BSS candidates are real BSSs, formed via mass transfer in binaries. This result is analogous to what found in Draco and Ursa Minor (paper I). 
We also note that BHB stars are less concentrated than the other considered populations (see $F_{\rm BSS,BHB}$), as already seen in Tolstoy et al. (2004). 

\subsection{Fornax}
The radial frequencies in Fornax are quite different  from those in Sculptor. In this galaxy the BSS candidates are more concentrated than all the other stellar populations taken into consideration (i.e. RHB, BHB, RGB, RC, young MS and BL). 

The fact that BSS candidates are more concentrated than all the other considered populations makes an identification with real BSSs  unlikely. 
BSS candidates are even more concentrated than BL stars (crosses in the right-hand panel of Fig.~\ref{fig:fig3}), which are usually associated with the young population of Fornax. This is likely due to the fact that
our selected BSS candidates, if they are genuine young stars, 
may contain stars between 0.2 and 1 Gyr, while the
selected BL stars are mostly between 0.4 and 0.7 Gyr. Indeed, B06
find that, when splitting the MS sample in stars approximately younger and
older than 0.4 Gyr on the basis of a color selection, then the spatial
distribution of the MS stars older than 0.4 Gyr agrees with the distribution of BL
stars, while the younger MS stars display a more concentrated distribution.

 Interestingly, the stellar population which most resembles the BSS candidates are the young MS stars. The radial distribution of young MS stars has the same trend as that of the BSS candidates within $\sim{}r_c$. Even at $r>r_c$ the relative frequency $F_{\rm BSS, MS}$ decreases less quickly than the other ones. This suggests that BSS candidates (or at least a large fraction of them) and young MS stars are the same population, as it was generally thought. This result strongly disfavours the association of most of BSS candidates in Fornax with real BSSs.
  Our sample of young MS stars might be contaminated by BHB stars, 
whose distribution can overlap that of the MS stars. This
contamination is hard to quantify, as the number and shape of the BHB
can change even in globular clusters of similar age and metallicity. 
However, we expect this contamination to be quite small, as only a
relatively weak ancient HB has been detected in Fornax (e.g., Stetson
et al. 1998; Bersier \&{} Wood 2002). It is clear from CMD analysis that
the ancient population of Fornax is only a small fraction of the 
total stellar population (e.g., Saviane et al. 2000).

  

\begin{table*}
\begin{center}
\caption{Number counts for Sculptor.}
\leavevmode
\begin{tabular}[!h]{lllllllll}
\hline
  $r\,{}$(arcsec)$^{\rm a}$
& $N_{{\rm BSS}}$$^{\rm b}$
& $\epsilon_{{\rm BSS}}$$^{\rm c}$
& $N_{{\rm RGB}}$$^{\rm b}$
& $\epsilon_{{\rm RGB}}$$^{\rm c}$
& $N_{{\rm RHB}}$$^{\rm b}$
& $\epsilon_{{\rm RHB}}$$^{\rm c}$
& $N_{{\rm BHB}}$$^{\rm b}$
& $\epsilon_{{\rm BHB}}$$^{\rm c}$\\
\hline
 198  & 158 (158) & 14.9 & 430 (435) & 21.0 & 235 (236)  & 15.7 & 171 (171)& 13.2\\
 522  & 119 (119) & 13.1 & 436 (438) & 21.1 & 181 (182)  & 13.7 & 176 (176)& 13.4\\
 800  & 129 (130) & 13.3 & 368 (378) & 19.6 & 115 (118)  & 11.2 & 215 (215)& 14.8\\
 1080 &  46 (47) &   8.4 & 191 (204) & 14.6 &  41  (45)  &  7.0 & 117 (117)& 10.9\\
 1450 &  39 (42) &   7.8 & 184 (219) & 15.7 &  28  (39)  &  6.8 & 142 (143)& 12.2\\
 2200 &  32 (40) &   8.5 & 220 (327) & 21.8 &  21  (55)  &  9.6 & 150 (153)& 12.6\\
 2850 &  16 (19) &   5.2 &  23  (66) & 10.0 &   0  (12)  &  4.5  & 26 (27) & 5.3\\
 3150 &   5  (8) &   4.0 &  18  (61) &  9.8 &   2  (16)  &  4.8  & 18 (19) & 4.4\\
\noalign{\vspace{0.1cm}}
\hline
\end{tabular}
\end{center}
\footnotesize{
$^{\rm {a}}$Centre of the annulus. $^{\rm {b}}$The value out of (in) the parenthesis is after (before) the subtraction of the foreground.  $^{\rm {c}}$ Poissonian error plus a term accounting for the uncertainty in the foreground subtraction and a term accounting for photometric errors (see Section~3).
}
\end{table*}
\begin{table*}
\begin{center}
\caption{Number counts for Fornax.}
 \leavevmode
\begin{tabular}[!h]{lllllllllll}
\hline
  $r\,{}$(arcsec)$^{\rm a}$
& $N_{{\rm BSS}}$$^{\rm b}$
& $\epsilon_{{\rm BSS}}$$^{\rm c}$
& $N_{{\rm RHB}}$$^{\rm b}$
& $\epsilon_{{\rm RHB}}$$^{\rm c}$
& $N_{{\rm BHB}}$$^{\rm b}$
& $\epsilon_{{\rm BHB}}$$^{\rm c}$
& $N_{{\rm RC}}$$^{\rm b}$
& $\epsilon_{{\rm RC}}$$^{\rm c}$
& $N_{{\rm BL}}$$^{\rm b}$
& $\epsilon_{{\rm BL}}$$^{\rm c}$
\\
\hline
 198  & 330 (331) & 20.7 &  154 (156)  & 15.2 & 42 (42)& 8.0 & 3145 (3153) & 57.7 & 245 (247) & 17.2\\
 522  & 372 (373) & 22.3 &  219 (221)  & 18.3 & 45 (45)& 8.4 & 4366 (4376) & 67.7 & 286 (289) & 18.7 \\
 800  & 286 (287) & 19.7 &  335 (339)  & 21.9 & 36 (37)& 7.8 & 5656 (5673)  & 76.9 & 226 (230) & 16.9\\
 1080 & 152 (153) & 14.5 &  264 (269)  & 19.7 & 43 (44)& 8.0 & 4323 (4343) & 67.4 & 189 (194) & 15.6 \\
 1450 & 140 (144) & 14.5 &  554 (568)  & 27.8 & 71 (73) & 10.5 & 5931 (5990) & 79.6 & 222 (237) & 17.8\\
 2200 &  66 (77) &  11.7 &  716 (758)  & 32.4  & 101 (106) & 12.5 & 5058 (5235) & 76.2 & 173 (217) & 18.1\\
 2850 &  12  (16) &  5.3 &  109 (126)  & 13.2 & 19 (21)  & 5.7 & 468 (541)   & 25.2 & 9 (27) & 6.7\\
 3150 &   4   (8) &  4.1 &  75 (91)    & 11.2  & 16 (18)  & 5.1 & 274 (340) & 20.4 & 5 (22) & 6.0\\
\noalign{\vspace{0.1cm}}
\hline
\end{tabular}
\end{center}
\footnotesize{
$^{\rm {a}}$Centre of the annulus. $^{\rm {b}}$The value out of (in) the parenthesis is after (before) the subtraction of the foreground.  $^{\rm {c}}$ Poissonian error plus a term accounting for  the uncertainty in the foreground subtraction and a term accounting for photometric errors (see Section~3).
}
\end{table*}
\begin{figure*}
\center{{ \epsfig{figure=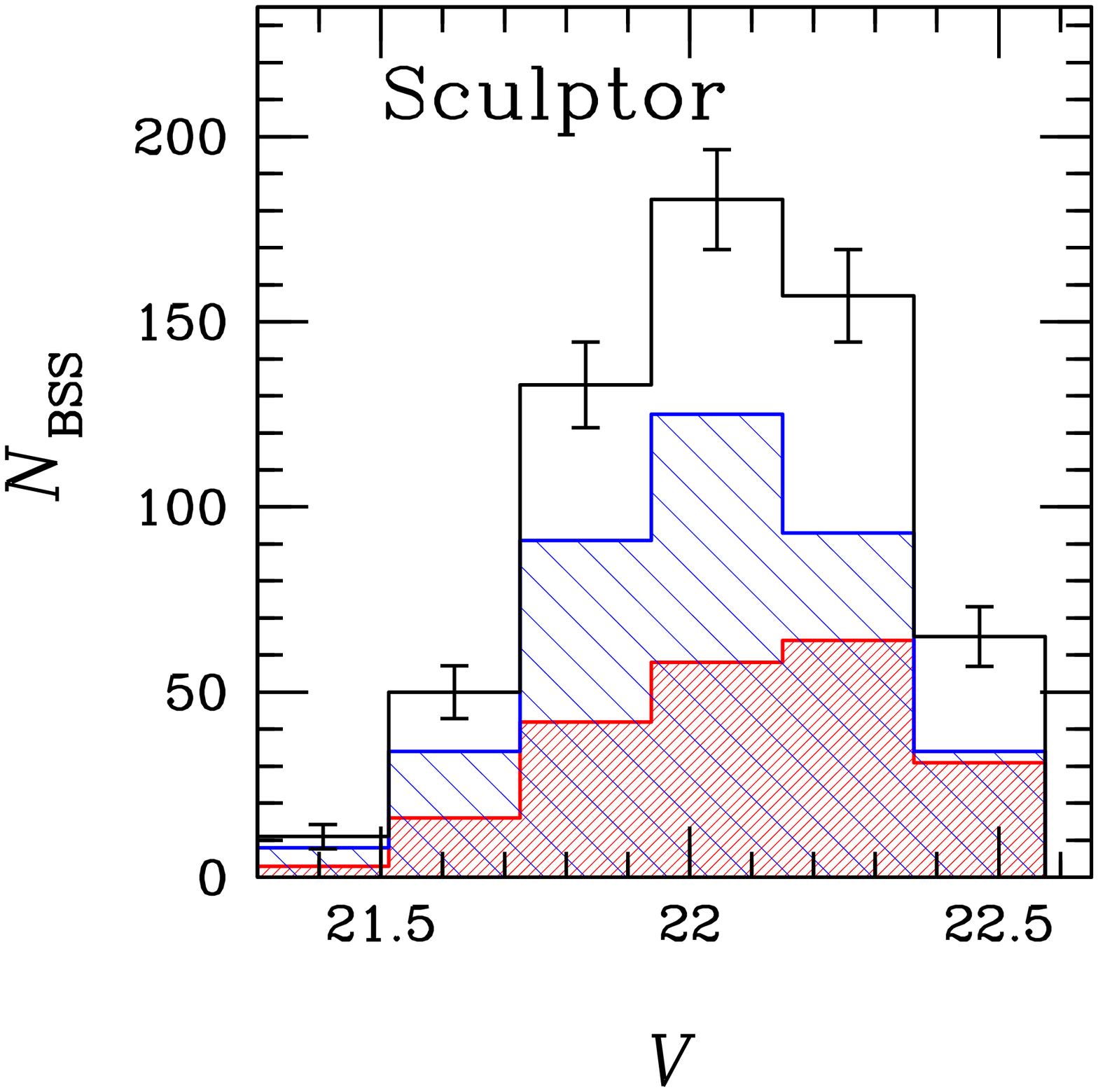,height=8cm}
\epsfig{figure=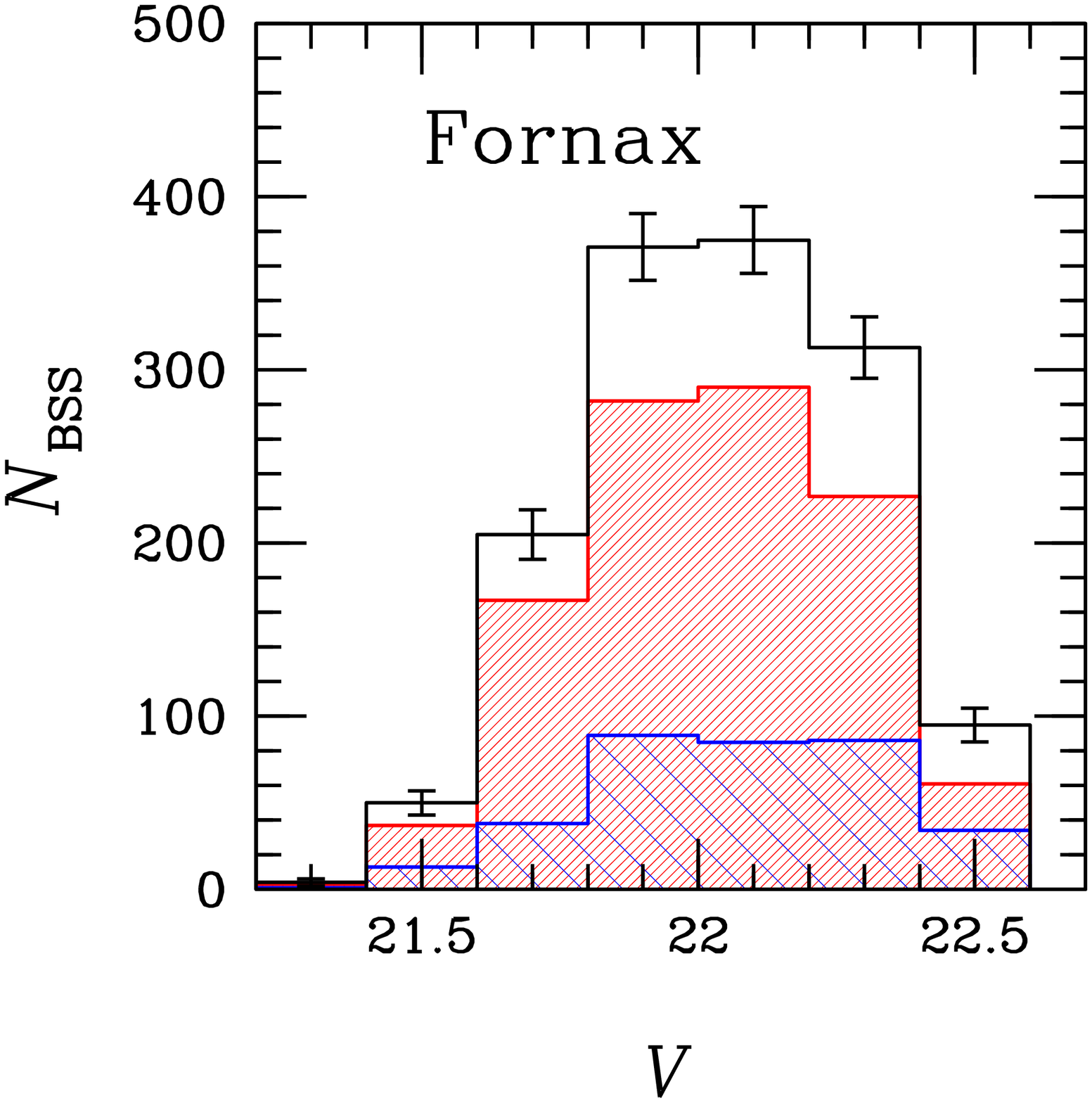,height=8cm} }}
\caption{\label{fig:fig4} Luminosity distribution of BSSs in Sculptor (left-hand panel) and Fornax (right-hand panel). The empty histogram represents the
entire sample of BSSs and the error bars show the Poissonian
errors. The lightly hatched (heavily hatched) histogram, blue (red) on the web, represents BSSs
with radial position $r>r_c$ ($r<r_c$).  }
\end{figure*}

\begin{figure*}
\center{{ \epsfig{figure=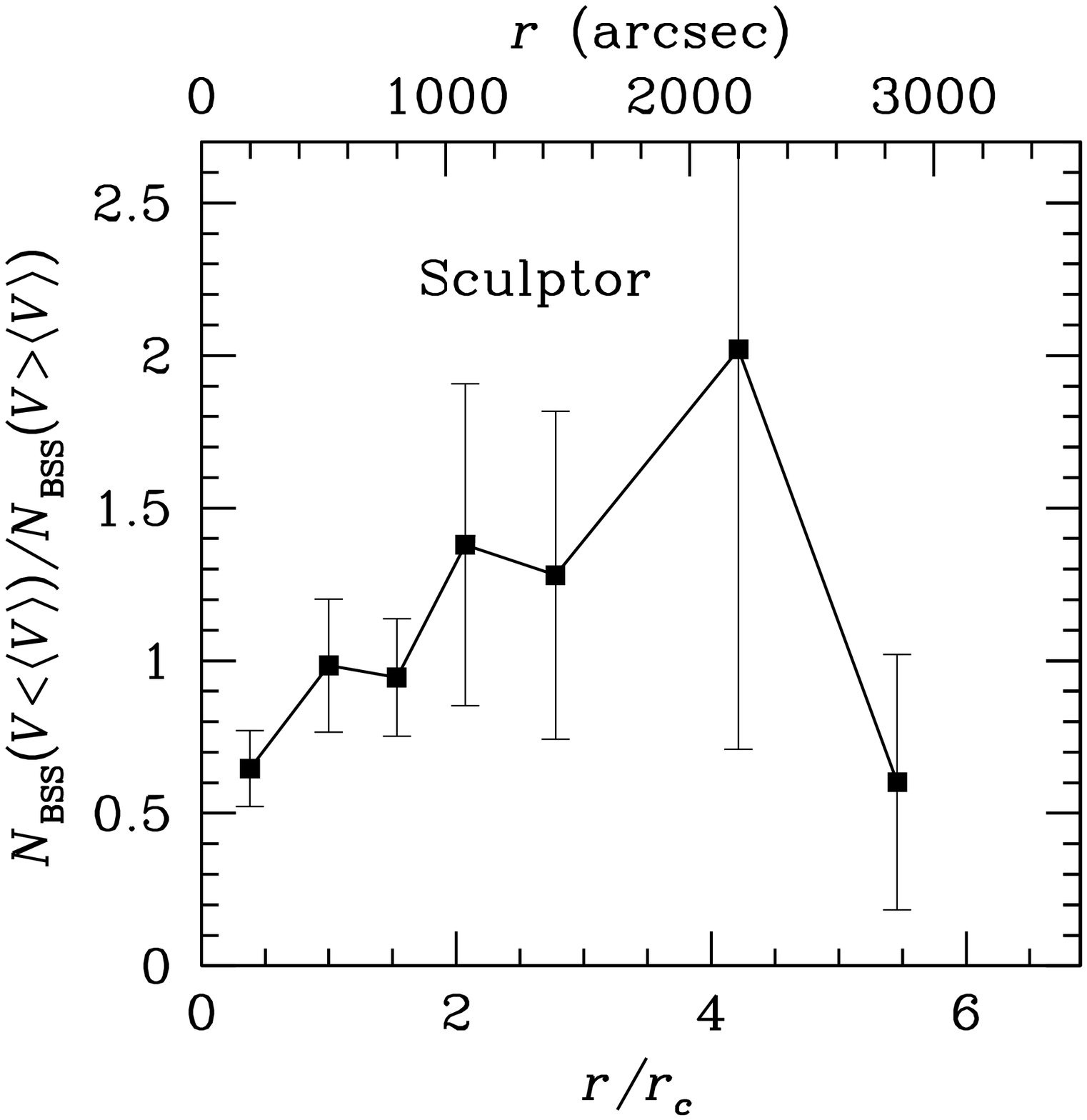,height=8cm} 
\epsfig{figure=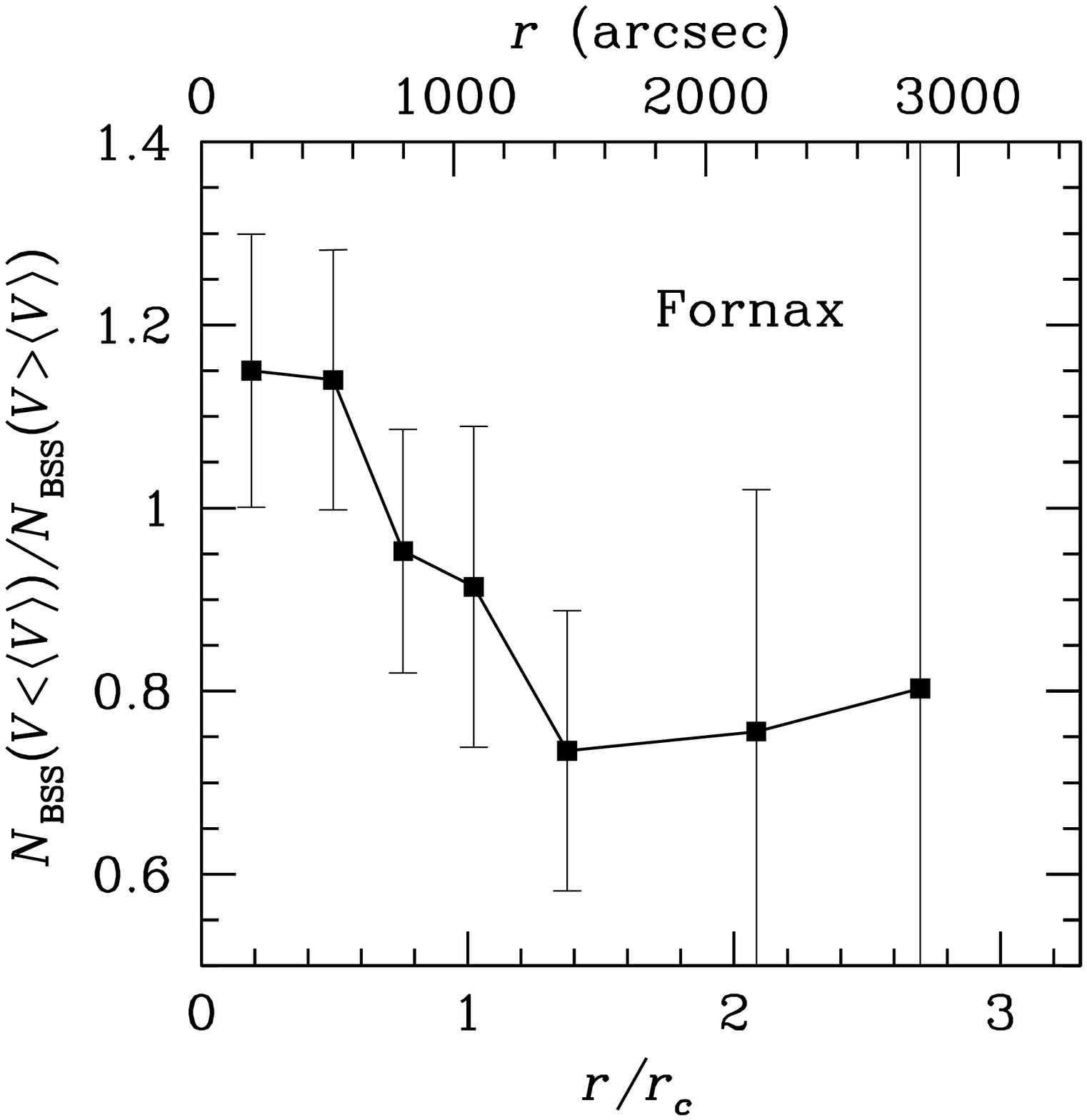,height=8cm} }} 
\caption{\label{fig:fig5} Relative frequency of bright BSSs versus faint BSSs in Sculptor (left-hand panel) and Fornax (right-hand panel). In both Sculptor and Fornax bright (faint) BSSs have been defined as the BSSs whose $V$ magnitude is lower (higher) than the average total value $\langle{}V\rangle{}$ ($\langle{}V\rangle{}=22.05$ for Sculptor and $\langle{}V\rangle{}=22.03$ for Fornax). The radial distributions have been corrected for foreground contamination. Error bars account for Poissonian statistics, uncertainties in foreground subtraction and photometric errors (see Section~3).}
\end{figure*}

\begin{figure*}
\center{{
\epsfig{figure=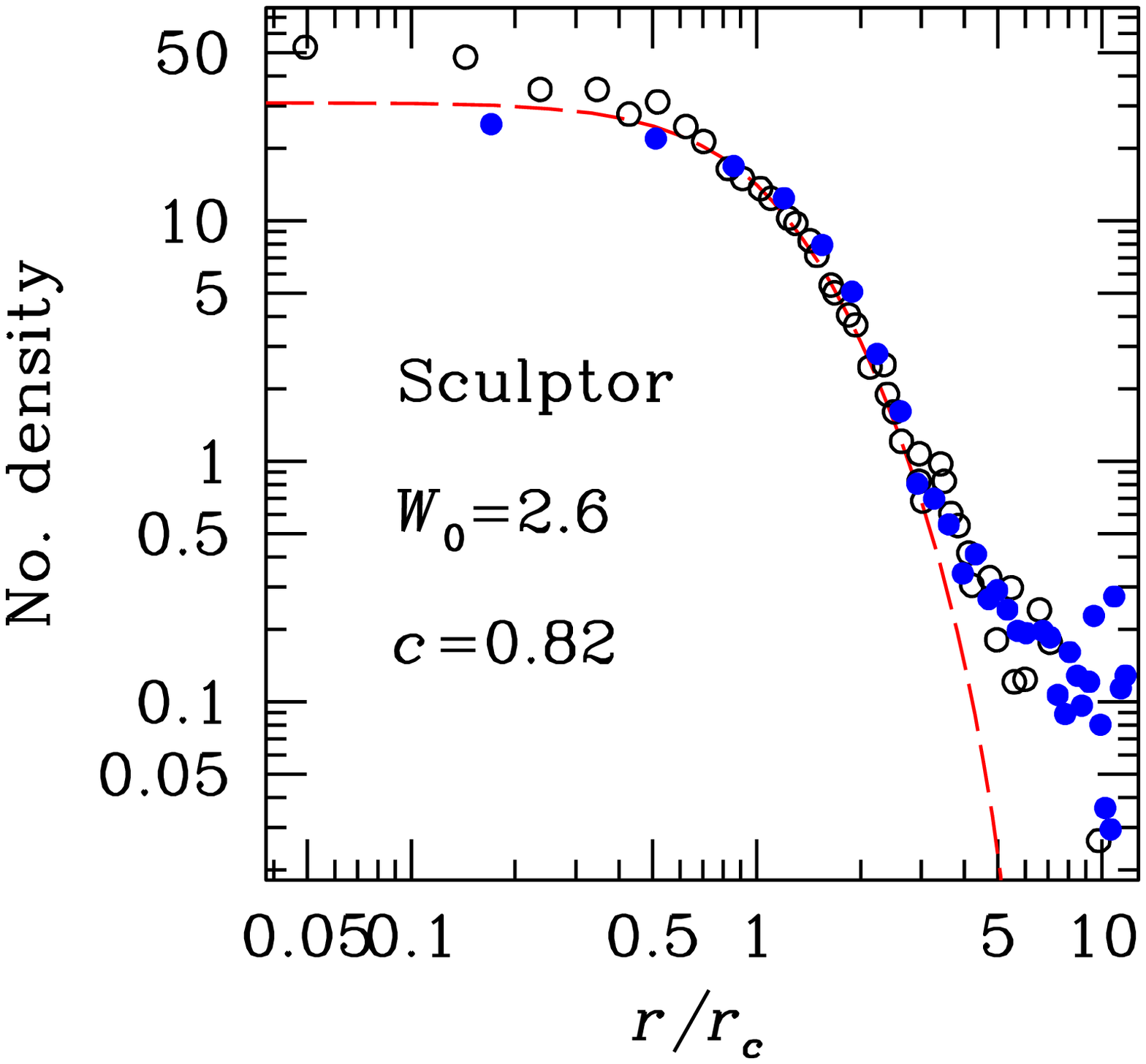,height=8cm} 
\epsfig{figure=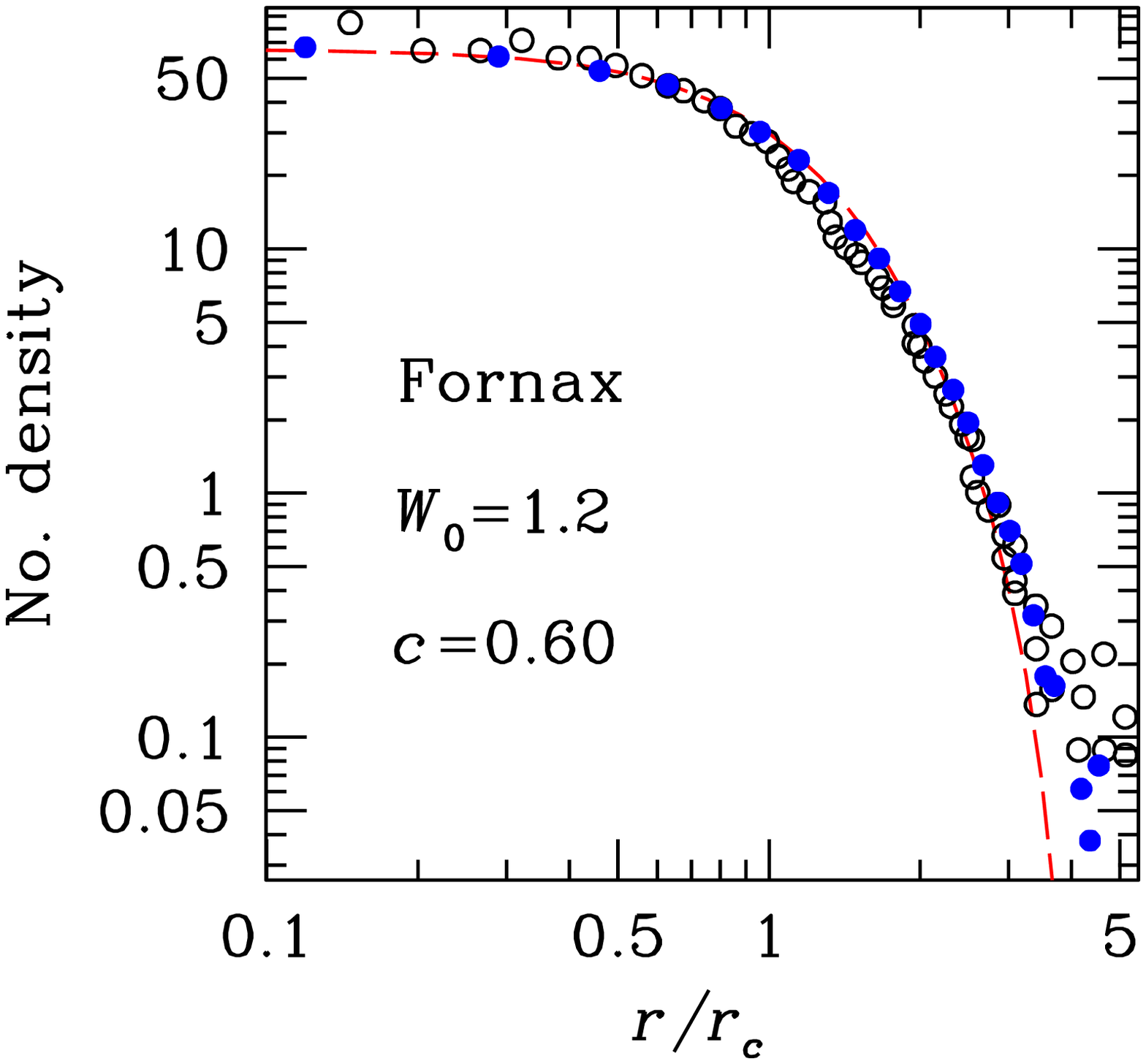,height=8cm} 
}}
\caption{\label{fig:fig6} Surface density profile of Sculptor (left-hand panel)
and Fornax (right-hand panel).  The number density is given in stars per squared arcmin. The open circles (black on the web) are data points from Irwin \&{} Hatzidimitriou 1995. The filled circles (blue on the web) are data points from B07 and from B06 for Sculptor and Fornax, respectively. The dashed
line (red on the web) is the best-fitting simulation. $r_c$ is 8.7' (B07) and 17.6' (B06) for Sculptor and Fornax, respectively.}
\end{figure*}

\section{Luminosity distribution}
The luminosity distribution is another interesting indicator of BSS properties in globular clusters. Observations suggest that in some globular clusters (Ferraro et al. 2003; Monkman et al. 2006) bright BSSs tend to be more concentrated than faint BSSs. Theoretical models indicate that collisional BSSs tend to be brighter than mass-transfer generated BSSs (Bailyn \&{} Pinsonneault 1995). No statistically significant correlation between position and luminosity has been found in Draco and Ursa Minor (paper~I), in agreement with the hypothesis that all the BSSs in these dSphs have been formed by mass-transfer in binaries.

 The left-hand panel of Fig.~\ref{fig:fig4} shows that, in Sculptor, BSSs which are outside $r_c$ (lightly hatched histogram) tend to be brighter than BSSs which are inside $r_c$ (heavily hatched histogram). 
On the contrary, in Fornax (right-hand panel of Fig.~\ref{fig:fig4}) BSSs which are outside $r_c$ (lightly hatched histogram) tend to be fainter than BSSs which are inside $r_c$ (heavily hatched histogram). 
Thus, there is a possible correlation between radial distance and luminosity of BSSs, and this correlation follows an opposite trend in Sculptor and in Fornax (i.e. it is a correlation in  Sculptor and an anti-correlation in Fornax). The trend observed in Fornax is similar to the one found in Sextans by Lee et al. (2003).
 
Fig.~\ref{fig:fig5} shows the radial distribution of bright BSSs normalized to the radial distribution of faint BSSs in Sculptor (left-hand panel) and Fornax (right-hand panel). In both Sculptor and Fornax bright (faint) BSSs have been defined as the BSSs whose $V$ magnitude is lower (higher) than the average total value $\langle{}V\rangle{}$ ($\langle{}V\rangle{}=22.05$ for Sculptor and $\langle{}V\rangle{}=22.03$ for Fornax). Fig.~\ref{fig:fig5} confirms that bright BSS candidates tend to be found preferentially in the inner parts in Fornax. Instead, the error bars for Sculptor in Fig.~\ref{fig:fig5} are quite large and the distribution is consistent with a flat one. Thus, in Fornax there is evidence of an anti-correlation between the luminosity and the radial distance of BSSs, whereas in Sculptor there is no statistically significant correlation.

\section{The simulations}

The data presented in the previous sections
show that BSS candidates in Sculptor behave approximately like mass-transfer BSSs, whereas the radial distribution of BSS candidates in Fornax is different from expectations for mass-transfer born BSSs, being more concentrated than that of RGB and RHB. 
In order to check the significance of this difference, we ran for Sculptor and Fornax the same kind of dynamical simulations that were performed in paper I for Draco and Ursa Minor and in previous papers (M04; M06) for BSSs in globular clusters.

\subsection{Method}
We adopt the upgraded version of the code by Sigurdsson \& Phinney
(1995) described in paper I (see also M04 and M06). 
The code integrates the
dynamics of BSSs, under the influence of the galactic potential, of
dynamical friction and of distant
encounters with other stars. Three-body encounters are implemented in
the code, but they are unimportant in the runs for dSphs. 
The potential of the host galaxy is represented by a time independent
multimass King model (see details in Sigurdsson \& Phinney 1995). 
To calculate the
potential, we input the observed core density ($n_c$) and velocity
dispersion ($\sigma{}_c$) of Sculptor and Fornax, and we modify the value of the
central adimensional potential, $W_0$ (defined in Sigurdsson \& Phinney
1995), until we reproduce the concentration and the
density profile of the galaxy under consideration. 
In Fig.~\ref{fig:fig6}
 the density profiles
of the best-fitting  King models are compared with the data. 

BSSs are generated with a given position, velocity and mass. Initial
positions are randomly chosen according to a probability distribution
homogeneous in the radial distance from the centre. This means that BSSs
are initially distributed according to an isothermal sphere, as we
expect for mass-transfer generated BSSs (see M04, M06). The minimum and the maximum value of
the distribution of initial radial distances, $r_{min}$ and $r_{max}$,
have been tuned in order to find the best-fitting simulation (Tables 4
and 5 report the most significant runs and their parameters for Sculptor
and Fornax, respectively).  In order to find the best fits, we consider also values of  $r_{max}$ smaller than the core radius $r_c$, although they are unphysical in those systems (such as dSphs) where collisions are unlikely and BSSs can form only via mass-transfer in binaries. In fact, mass-transfer generated BSSs should track the distribution of the progenitor binaries. Only collisionally generated BSSs are expected to be concentrated in the core.
 
 For Fornax we made some check runs where an initial  offset $r_{off}$ up to
0.1 $r_c$ is given to the centre-of-mass of BSS distribution, in order to match the observed offset (see Section~6) of BSS candidates in Fornax with respect to the other stellar populations. However, the values of the non-reduced $\chi{}^2$ for the runs with the offset changes by less than
15 per cent with respect to the same runs without offset (see the runs Fnx16 and Fnx17 in Table~4).
Initial velocities are generated from the distributions described in
Sigurdsson \& Phinney (1995). No initial kicks
 are given to BSSs,  because  they are expected to be
mass-transfer BSSs. 


\begin{table*}
\begin{center}
\caption{Simulation parameters and $\chi{}^2$ for Sculptor.}
\leavevmode
\begin{tabular}[!h]{llllllll}
\hline
Run 
&  $r_{min}/r_c$
& $r_{max}/r_c$
& $t_{last}$ (Gyr)
& $m_{\rm BSS}$ ($M_\odot{}$)
& $\chi{}_{\rm RGB}^2$
& $\chi{}_{\rm RHB}^2$
& $\chi{}_{\rm BHB}^2$\\
\hline

 Scl1       & 0.0 & 2.0 & 2 & 1.3  &  28  & 17  & 25 \\
 Scl2       & 0.0 & 2.9 & 2 & 1.3  &  6.7 & 4.0 & 6.1\\
 Scl3       & 0.0 & 3.0 & 2 & 1.3  &  6.6 & 3.8 & 6.3\\
 Scl4       & 0.0 & 4.0 & 2 & 1.3  &  21  & 12 & 19\\ 
 Scl5       & 0.0 & 5.0 & 2 & 1.3  &  53  & 30 & 48\\

 Scl6       & 0.1 & 2.9 & 2 & 1.3  &  6.7 & 3.7 & 6.3\\
 Scl7       & 0.1 & 4.0 & 2 & 1.3  &  27  & 16 & 24\\

 {\bf Scl8} & 0.2 & 2.9 & 2 & 1.3  &  {\bf 5.0}  & {\bf 2.8} & {\bf 4.7}\\
 
 Scl9       & 0.2 & 3.0 & 2 & 1.3  &  7.6  & 4.4  & 7.2\\

 Scl10      & 0.3 & 2.9 & 2 & 1.3  &  6.8 &  4.1 &  6.3\\

 Scl11       & 0.6 & 2.9 & 2 & 1.3  &  12  & 7.8 & 10\\
 Scl12       & 0.2 & 2.9 & 1 & 1.3  &  6.1 & 3.6 & 5.6\\
 Scl13       & 0.2 & 2.9 & 4 & 1.3  &  6.0 & 3.5 & 5.7\\
 Scl14       & 0.2 & 2.9 & 10 & 1.3  & 6.5  & 3.6 & 6.2 \\
 Scl15       & 0.2 & 2.9 & 2 & 1.1  & 6.7 & 4.0 & 6.2\\
 Scl16       & 0.2 & 2.9 & 2 & 1.5  & 5.6 & 3.2 & 5.3\\


\noalign{\vspace{0.1cm}}
\hline
\end{tabular}
\end{center}
\footnotesize{
$r_{min}$ and $r_{max}$ are the minimum and maximum distance, 
$t_{last}$ and $m_{\rm BSS}$  are the maximum lifetime and the mass of a BSS.\\
$\chi{}_{\rm RGB}^2$, $\chi{}_{\rm RHB}^2$ and  $\chi{}_{\rm BHB}^2$ indicate the $\chi{}^2$ of $F_{\rm BSS,RGB}$, $F_{\rm BSS,RHB}$ and $F_{\rm BSS,BHB}$, respectively. The reported values of $\chi{}_{\rm RGB}^2$,  $\chi{}_{\rm RHB}^2$ and  $\chi{}_{\rm BHB}^2$  are not reduced and have been calculated on the basis of 6 data points. The best-fitting model is indicated in bold face.
}
\end{table*}

\begin{table*}
\begin{center}
\caption{Simulation parameters and $\chi{}^2$ for Fornax.}
\leavevmode
\begin{tabular}[!h]{lllllllll}
\hline
Run 
&  $r_{min}/r_c$
& $r_{max}/r_c$
& $t_{last}$ (Gyr)
& $m_{\rm BSS}$ ($M_\odot{}$)
& $r_{off}/r_c$
& $\chi{}_{\rm RGB}^2$
& $\chi{}_{\rm RHB}^2$
& $\chi{}_{\rm BHB}^2$\\
\hline
 Fnx1       & 0.0 & 0.5 & 2 & 2.0 & 0 &  37  & 17 & 5.7\\

 {\bf Fnx2} & 0.0 & 0.7 & 2 & 2.0 & 0  &  {\bf 13}  & {\bf 7.1} & {\bf 2.9}\\
 
 Fnx3       & 0.0 & 1.0 & 2 & 2.0 & 0  &  14  & 8.2 & 3.3\\ 
 Fnx4       & 0.0 & 1.2 & 2 & 2.0 & 0  &  27  & 15 & 5.5\\ 
 Fnx5       & 0.0 & 1.3 & 2 & 2.0 & 0  &  45 &  24 & 8.9\\
 Fnx6       & 0.0 & 1.5 & 2 & 2.0 & 0  &  72  & 39 & 14\\ 
 Fnx7       & 0.2 & 1.0 & 2 & 2.0 & 0  &  23  & 12 & 4.8\\
 Fnx8       & 0.3 & 1.0 & 2 & 2.0 & 0  &  34  & 17 & 6.5\\
 Fnx9       & 0.4 & 1.0 & 2 & 2.0 & 0  &  37  & 18 & 6.6\\
 Fnx10      & 0.2 & 1.5 & 2 & 2.0 & 0  &  106  & 55 & 19\\
 Fnx11      & 0.0 & 0.7 & 1 & 2.0 & 0  &  20  & 11 & 4.5\\
 Fnx12      & 0.0 & 0.7 & 4 & 2.0  & 0 &  15  & 7.8 & 3.1\\
 Fnx13      & 0.0 & 0.7 & 10 & 2.0 & 0  &  17  & 9.3 & 3.8\\
 Fnx14      & 0.0 & 0.7 & 2 & 1.8  &  0 & 13  & 7.1 & 3.0\\
 Fnx15      & 0.0 & 0.7 & 2 & 2.3  &  0 & 13  & 7.1 & 3.0\\
 Fnx16      & 0.0 & 0.7 & 2 & 2.0  &  0.05 & 15  & 8.0 & 3.3\\
 Fnx17      & 0.0 & 0.7 & 2 & 2.0  &  0.1 & 13  & 7.4 &  3.4\\
\noalign{\vspace{0.1cm}}
\hline
\end{tabular}
\end{center}
\footnotesize{
$r_{min}$ and $r_{max}$ are the minimum and maximum distance,
$t_{last}$ and $m_{\rm BSS}$  are the maximum lifetime and the mass of a BSS. $r_{off}$ is the offset of BSS centre-of-mass with respect to Fornax centre-of-mass. \\
$\chi{}_{\rm RGB}^2$, $\chi{}_{\rm RHB}^2$ and  $\chi{}_{\rm BHB}^2$ indicate the $\chi{}^2$ of $F_{\rm BSS,RGB}$, $F_{\rm BSS,RHB}$ and $F_{\rm BSS,BHB}$, respectively. The reported values of $\chi{}_{\rm RGB}^2$,  $\chi{}_{\rm RHB}^2$ and  $\chi{}_{\rm BHB}^2$  are not reduced and have been calculated on the basis of 8 data points (7 data points in the case of  $\chi{}_{\rm BHB}^2$).  The best-fitting model is indicated in bold face.
}
\end{table*}

 In most of the runs the mass of the BSSs is assumed to be $m_{\rm
BSS}=1.3\,{}M_\odot{}$ for Sculptor and $m_{\rm
BSS}=2.0\,{}M_\odot{}$ for Fornax, as indicated by the isochrones  for our data  (see the Appendix~A). We made check runs with masses in the range from
1.1 to 1.5 $M_\odot{}$ and from 1.8 to 2.3 $M_\odot{}$ for Sculptor and Fornax, respectively. This range of masses
is also consistent with the isochrones (see Appendix A).

Each BSS is evolved for a time $t$, randomly selected  from a
homogeneous distribution between $t=0$ and $t=t_{last}$. The parameter
$t_{last}$ represents the lifetime of BSSs (see M04, M06). We made runs
with $t_{last}$=1, 2, 4, 10 Gyr (see paper~I for details about these choices).


\subsection{Comparison with observations}
\begin{figure*}
\center{{
\epsfig{figure=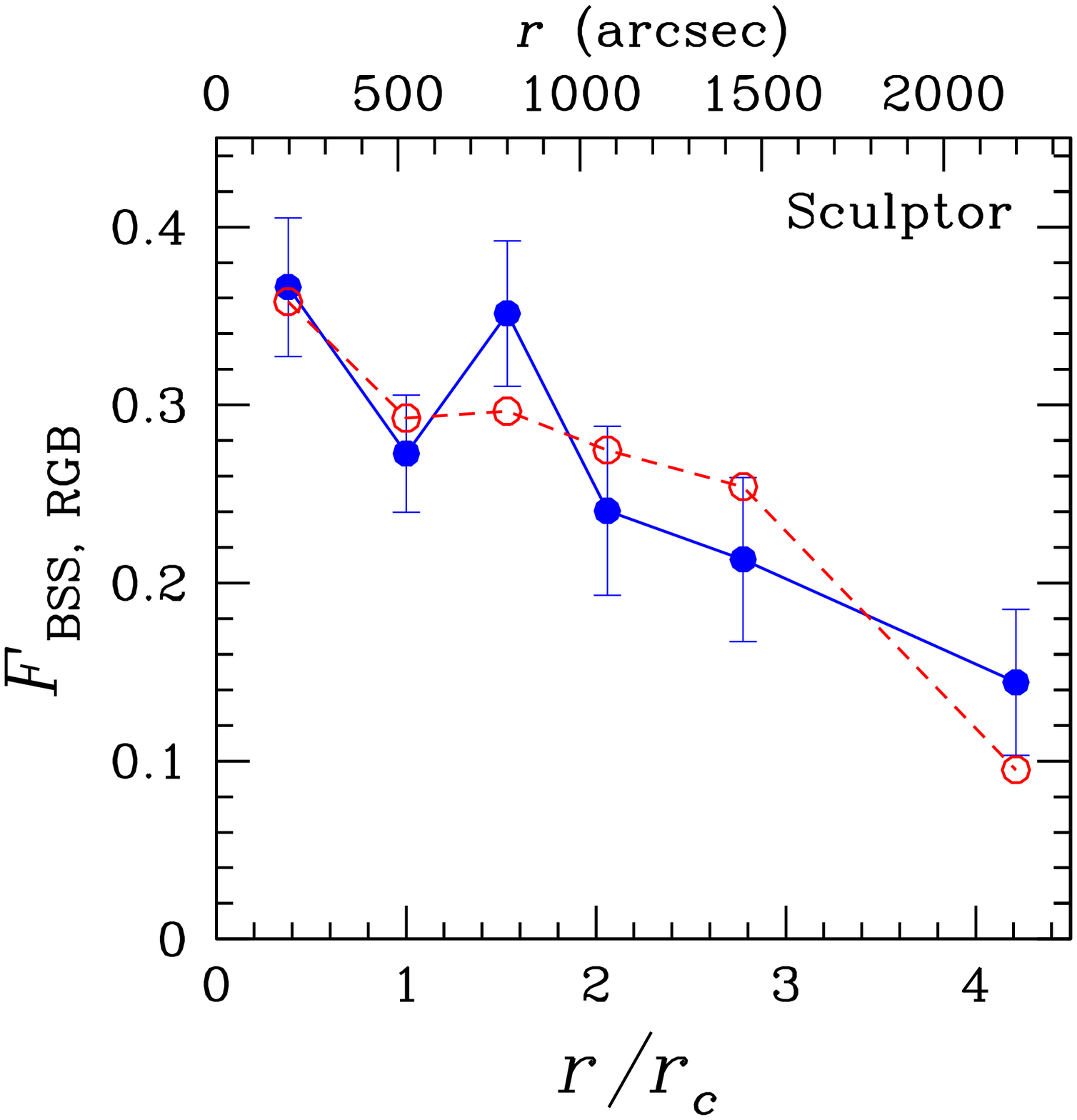,height=5cm} 
\epsfig{figure=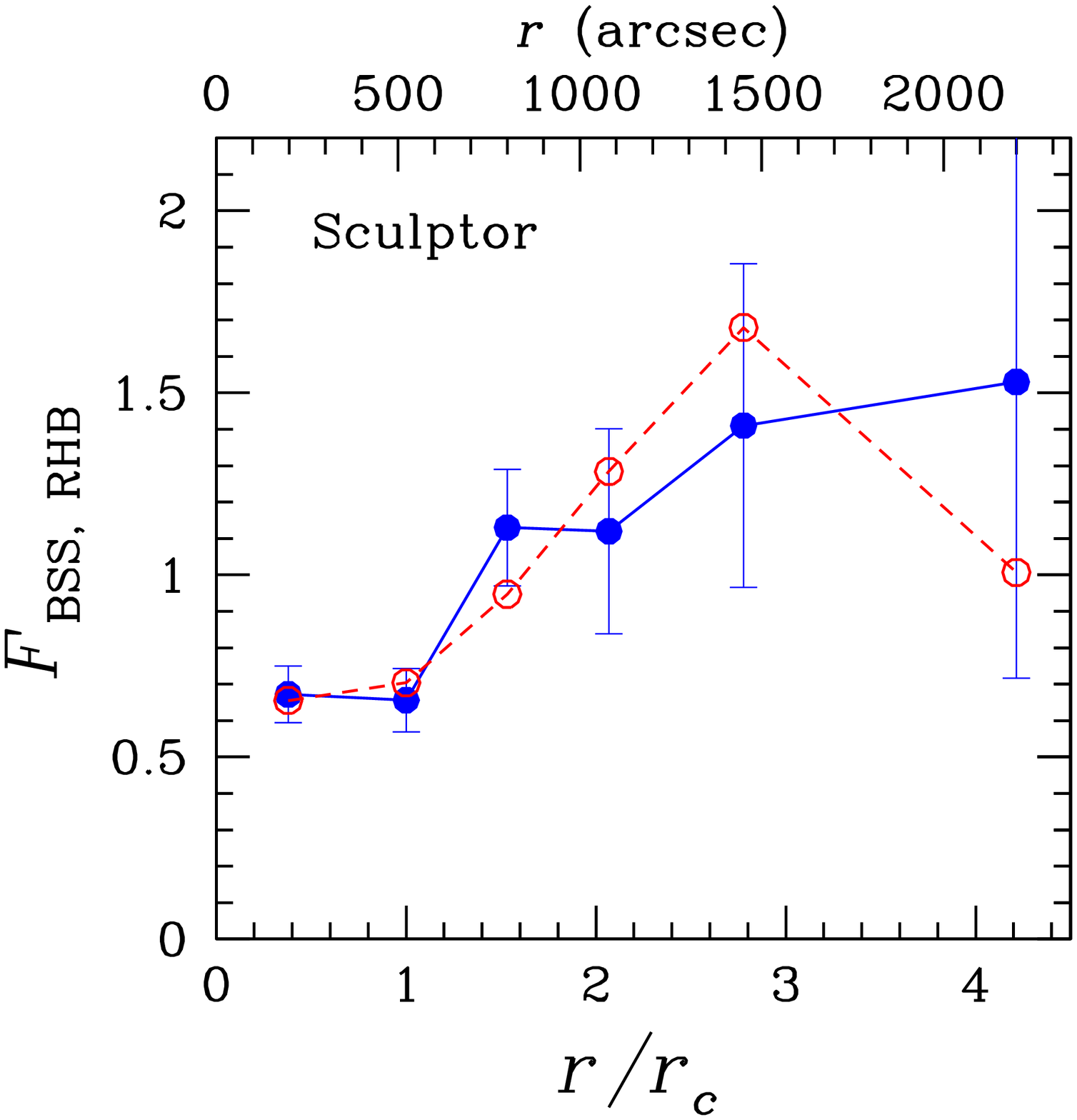,height=5cm} 
\epsfig{figure=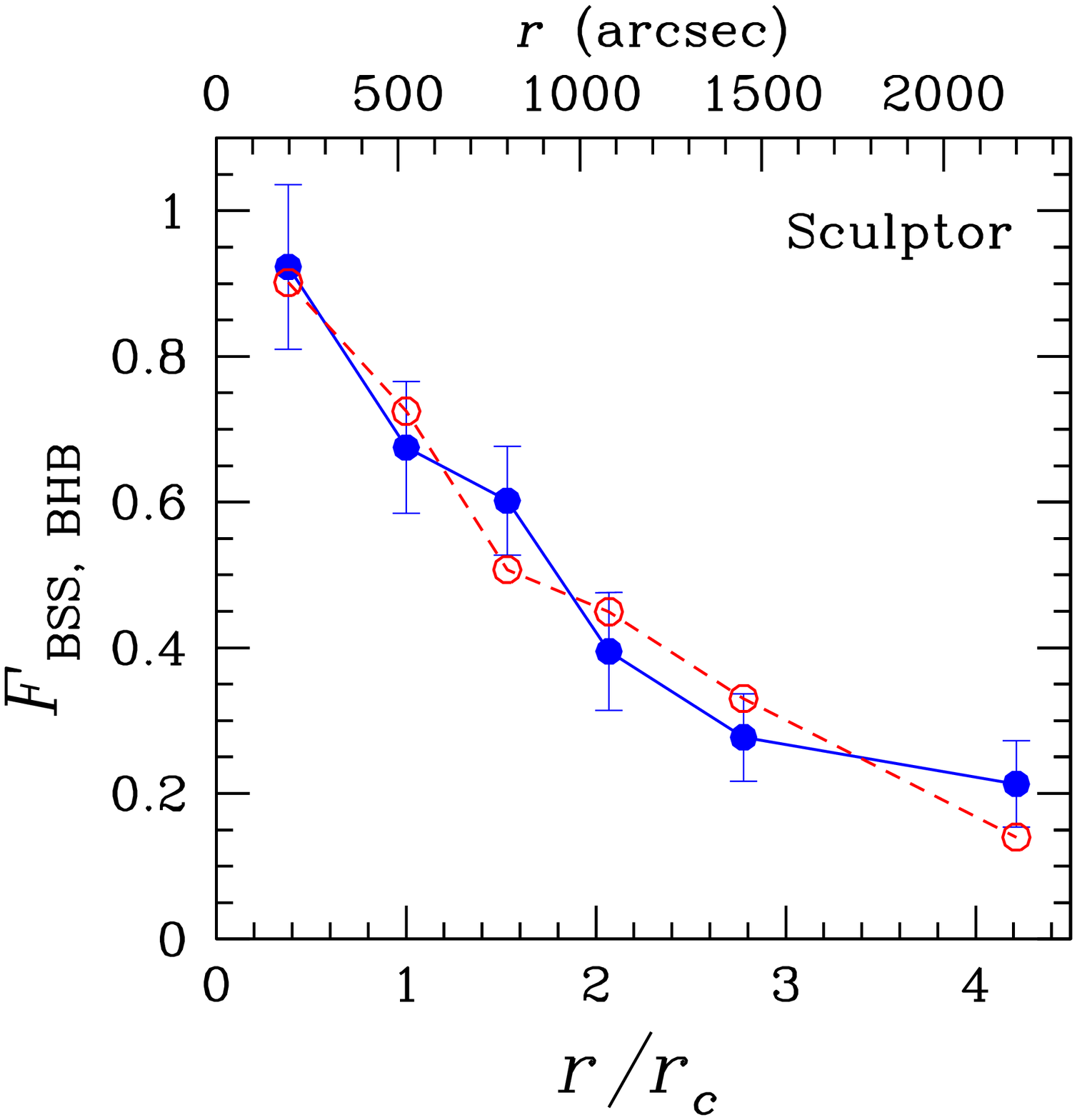,height=5cm} 
}}
\caption{\label{fig:fig7}
Left-hand panel: relative frequency of BSSs normalized to RGB in Sculptor. 
Central panel: relative frequency of BSSs normalized to RHB in Sculptor. 
Right-hand panel:  relative frequency of BSSs normalized to BHB in Sculptor. 
In all panels the filled circles connected by the solid line, blue on the web, are the observations with the corresponding error bars (the same as in Fig.~\ref{fig:fig3}). The open circles connected by the dashed line, red on the web, are the best-fitting model (Scl8).
}
\end{figure*}
\begin{figure*}
\center{{
\epsfig{figure=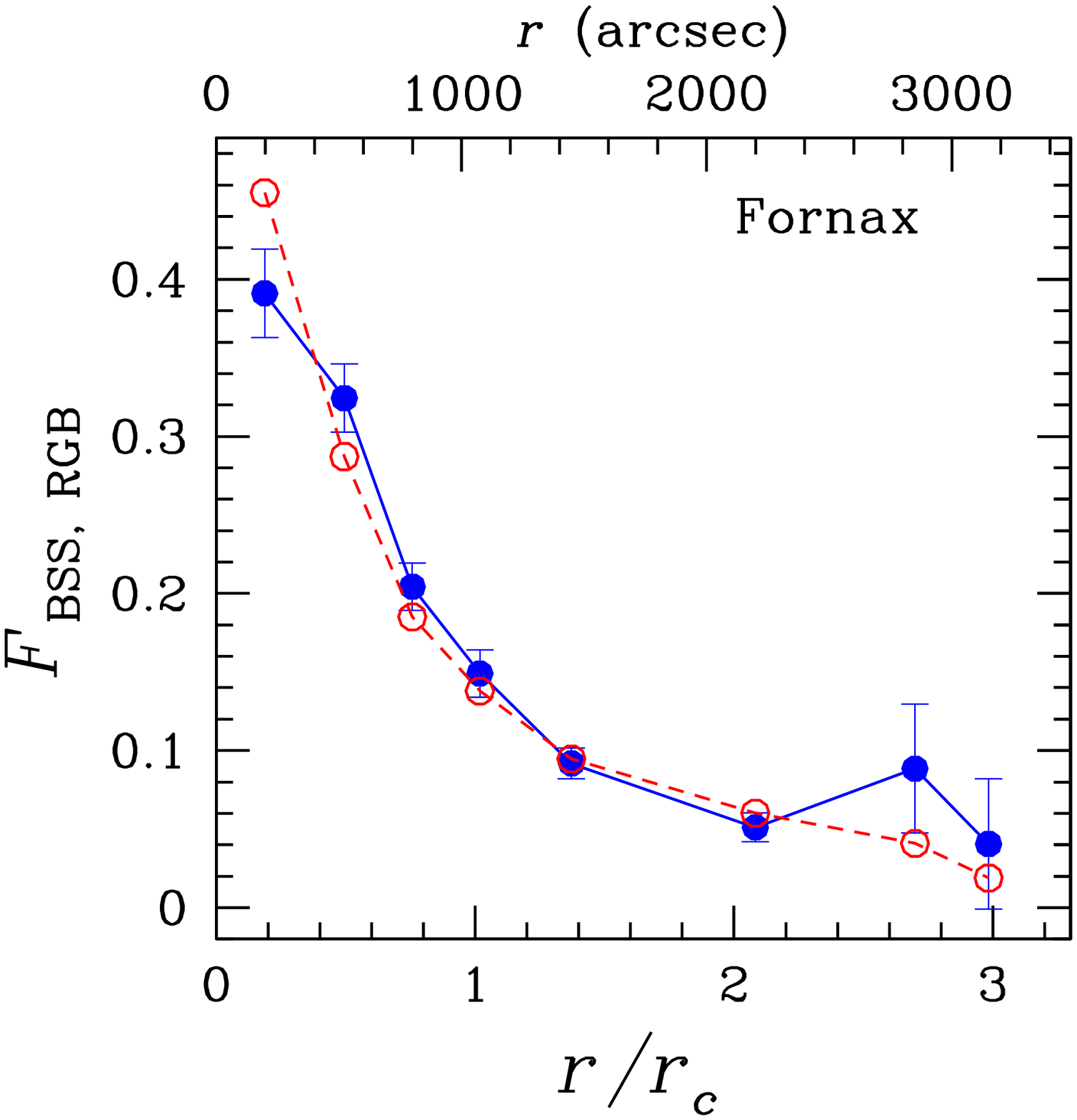,height=5cm} 
\epsfig{figure=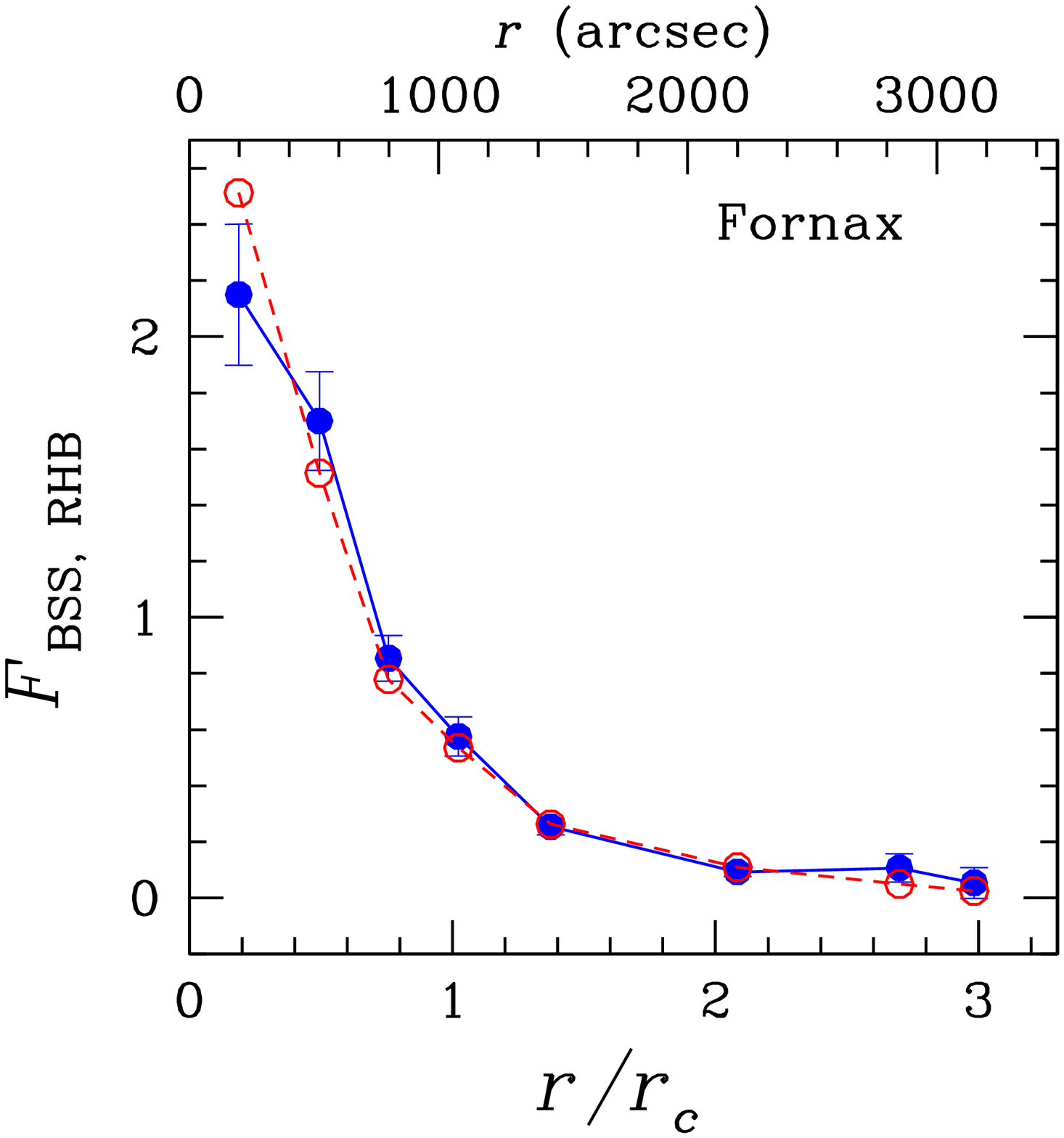,height=5cm} 
\epsfig{figure=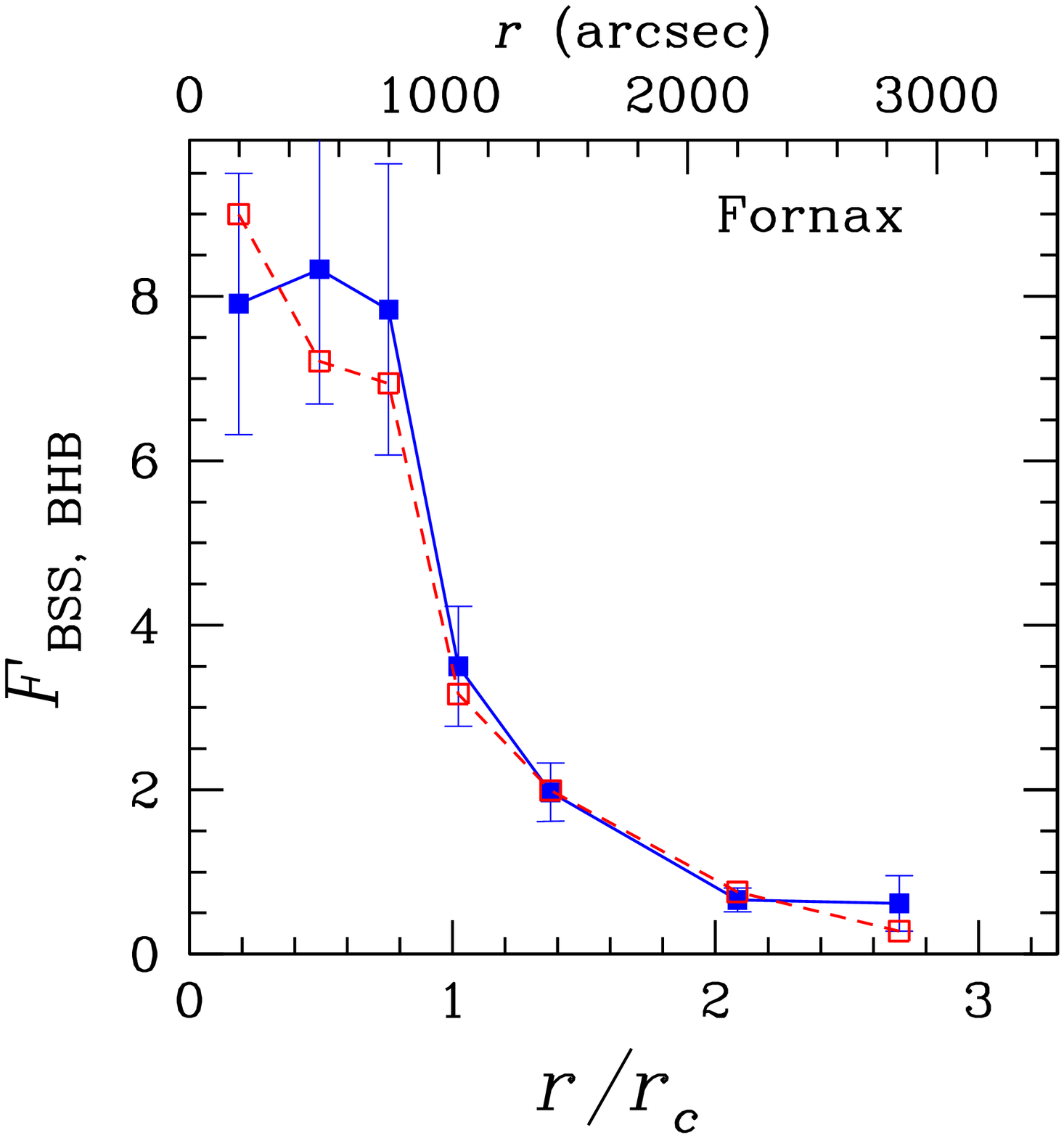,height=5cm} 
}}
\caption{\label{fig:fig8}
Left-hand panel: relative frequency of BSSs normalized to RGB in Fornax. 
Central panel: relative frequency of BSSs normalized to RHB in Fornax. 
Right-hand panel:  relative frequency of BSSs normalized to BHB in Fornax. 
In all panels the filled circles connected by the solid line, blue on the web, are the observations  with the corresponding error bars (the same as in Fig.~\ref{fig:fig3}). The open circles connected by the dashed line, red on the web, are the best-fitting model (Fnx2).
}
\end{figure*}
\subsubsection{Sculptor}
  In the case of Sculptor the best-fit is achieved for $r_{min}=0.2\,{}r_c$ and $r_{max}=2.9\,{}r_c$ (case Scl8 in Table~4; see Fig.~\ref{fig:fig7}). However, all the models with $r_{min}\sim{}0-0.3\,{}r_c$ and $r_{max}\sim{}3\,{}r_c$ have acceptably low values of $\chi{}^2$ (i.e. non-reduced $\chi{}^2<10$, with 6 data points). 

The best matching simulation for Sculptor has  $\chi{}^2_{\rm RGB}\sim{}5$, $\chi{}^2_{\rm RHB}\sim{}3$ and $\chi{}^2_{\rm BHB}\sim{}5$. These values refer to non-reduced $\chi{}^2$ with 6 data points. The corresponding values of the reduced $\chi{}^2$ (considering that there are 2 main parameters, i.e. $r_{min}$ and $r_{max}$) are $\tilde{\chi{}}^2_{\rm RGB}\sim{}1.2$, $\tilde{\chi{}}^2_{\rm RHB}\sim{}0.7$ and $\tilde{\chi{}}^2_{\rm BHB}\sim{}1.2$. 
Thus, the best matching simulation for Sculptor (Scl8 in Table~4) is in reasonable agreement with the data, and the observed BSS candidates behave like mass-transfer generated BSSs, although new data with smaller photometric errors and a deeper completeness limit are required, in order to  obtain a more accurate radial distribution.


\subsubsection{Fornax}

 In the case of Fornax the $\chi{}^2$ of the best-fitting simulation  is quite higher than for Sculptor. For the best-matching run (Fnx2 in Table~5, with $r_{min}=0.0\,{}r_c$ and $r_{max}=0.7\,{}r_c$; see Fig.~\ref{fig:fig8}) the values of non-reduced $\chi{}^2$ are $\chi{}^2_{\rm RGB}\sim{}13$, $\chi{}^2_{\rm RHB}\sim{}7$ and $\chi{}^2_{\rm BHB}\sim{}3$, corresponding to reduced $\chi{}^2$ values $\tilde{\chi{}}^2_{\rm RGB}\sim{}2.2$, $\tilde{\chi{}}^2_{\rm RHB}\sim{}1.2$ and $\tilde{\chi{}}^2_{\rm BHB}\sim{}0.6$. 
Thus, the relative frequency $F_{\rm BSS, RGB}$ obtained from the best-matching simulation Fnx2 is only marginally in agreement with the observed distribution, whereas the simulated values of $F_{\rm BSS, RHB}$ and $F_{\rm BSS, BHB}$ are consistent with the observations.

The most interesting result about Fornax is that only runs with $r_{max}\lesssim{}1.0\,{}r_c$ have non-reduced $\chi{}_{\rm RGB}^2<20$, implying that most of BSS candidates in Fornax are concentrated inside the core. This result is not only very different from Draco, Ursa Minor and Sculptor, but also hard to explain within the mass-transfer model for the formation of BSSs. In fact, mass-transfer generated BSSs should exist throughout the host galaxy, as they  simply track the distribution of binaries. There is no reason why BSSs should be present almost exclusively in the core, unless the core itself is a collisional environment, where BSSs may form through stellar collisions (which is not the case of Fornax). 

If we still assume that BSS candidates in Fornax are real BSSs, a possible explanation for the concentration of BSSs in the core might be that BSSs in the core of Fornax have been ejected from globular clusters (either the existing ones, or ancient clusters which are now disrupted). 
However, this interpretation does not seem viable, for the following reasons.

\begin{figure*}
\center{{
\epsfig{figure=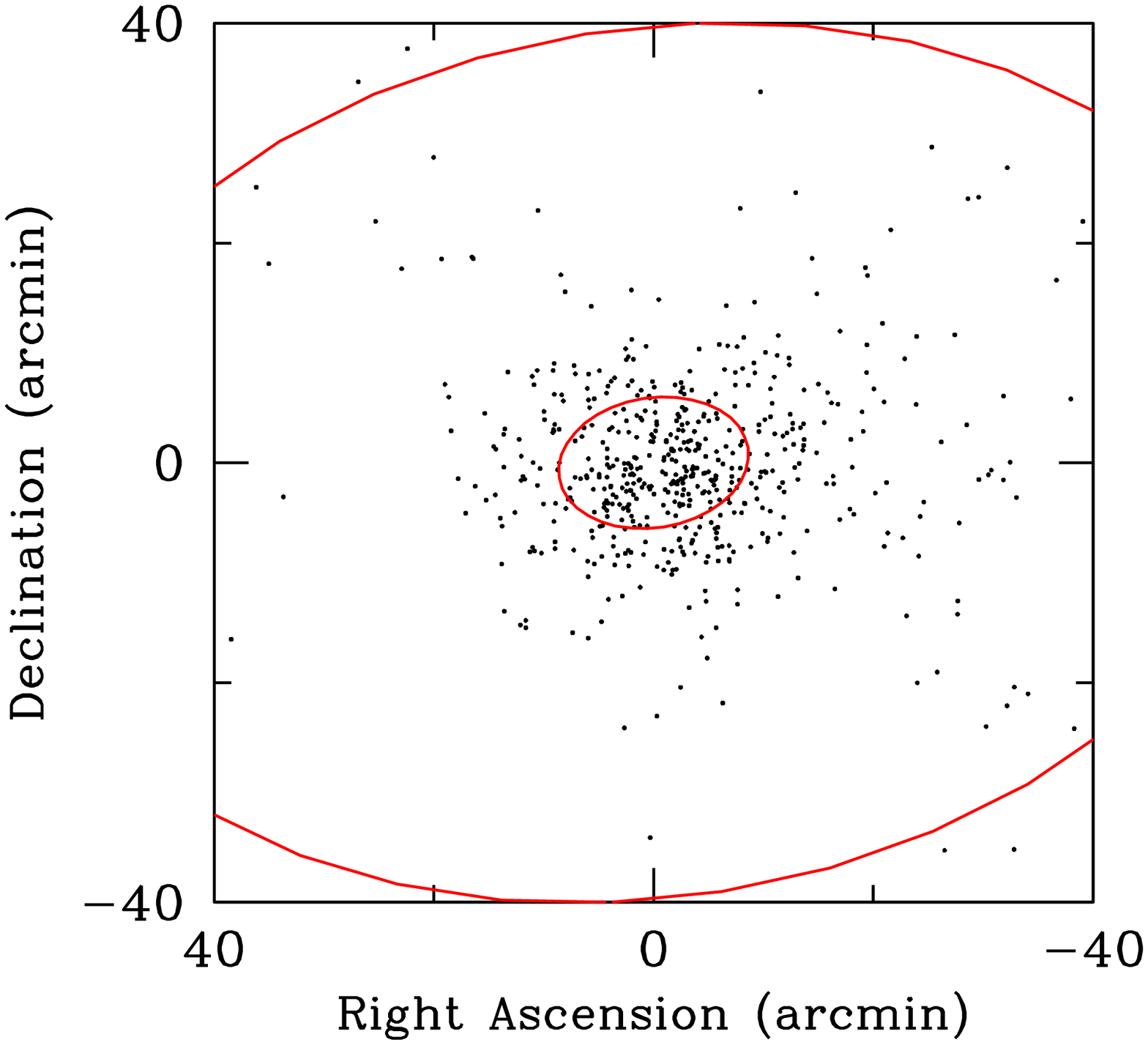,height=8.0cm} 
\epsfig{figure=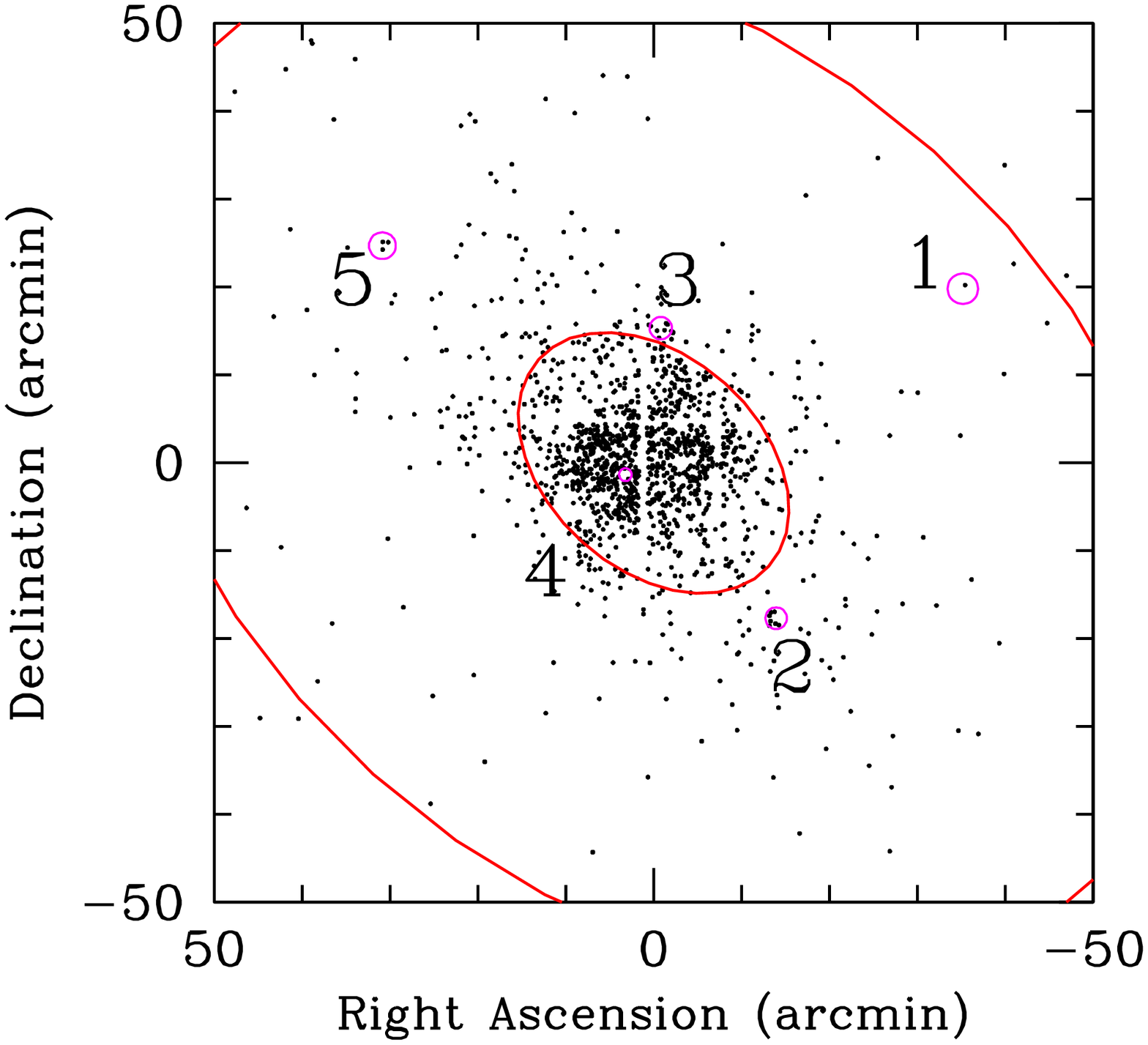,height=8.0cm}

}}
\caption{\label{fig:fig9}
Right Ascension and Declination of the BSS candidates imaged in Sculptor (left-hand panel) and in Fornax (right-hand panel). The concentric ellipses (red on the web) indicate tidal and core radii ($r_t$
and $r_c$; same as Fig.~\ref{fig:fig1}). The small circles (magenta on the web) indicate the area within the tidal radii of the globular clusters in Fornax. The globular clusters are labeled with numbers from 1 to 5 (see B06).
}
\end{figure*}

The number of observed BSSs within the core of Fornax is quite high ($\sim{}1067$, 76 per cent of the entire BSS population). For comparison, the Galactic globular cluster with the largest observed number of BSSs, $\omega{}$ Centauri, has $\sim{}300$ observed BSSs (Ferraro et al. 2006). 
Only 1 of the 5 globular clusters currently present in Fornax is within $r_c$ (right-panel of Fig.~\ref{fig:fig9}).
Thus, in order to match the inner component of the BSS population, not only the innermost globular cluster of Fornax should host an uncommonly large population of BSSs, but a large fraction of them need to be ejected from the cluster. 
An alternative hypothesis is that  in the past the core of Fornax hosted other globular clusters, which have completely disappeared due to tidal stripping. This would allow a high fraction of BSSs to be deposited in the core. 
However, an improbably large number of massive globular clusters (i.e. about $3-4$ globular clusters with the same BSS content of $\omega{}$ Centauri)  need to be tidally stripped, in order to produce all the BSSs observed in the core of Fornax.

Thus, the high $\chi{}^2$ obtained from the simulations and especially the fact that BSS candidates are too concentrated in the core of the host galaxy to be mass-transfer generated BSSs indicate that BSS candidates in Fornax (or at least a large fraction of them) are not real BSSs. 
 Instead, a mixed scenario in which most of BSS candidates in Fornax are young MS stars and only a small fraction of them are real BSSs is favoured from both the data and  the simulations.
 In fact, the very concentrated radial distribution of BSS candidates is consistent with the fact that in many dwarf galaxies younger stars appear more centrally concentrated than the older ones (Baade \&{} Gaposchkin 1963; Skillman et al. 2003). This is likely due to the fact that  gas was retained for a longer time in the inner parts, where the potential is deeper.
Furthermore, the brighter BSS candidates, concentrated in the inner part of the galaxy, may be connected with young stars, whereas the fainter ones might be real BSSs, accounting for the trend in the luminosity distribution (right-hand panel of Fig.~\ref{fig:fig5}). 
Finally, the scenario in which most BSS candidates in Fornax are young MS stars is indirectly supported by the fact that there are clear signs of a young population throughout the CMD of Fornax (e.g. the existence of  BL stars with an age of 0.4-0.7 Gyr, see B06 and the Appendix).



\section{Asymmetry in the BSS distribution}
Fig.~\ref{fig:fig9} contains other interesting information about the spatial distribution of BSS candidates: the distribution of BSS candidates in Fornax (right-hand panel of Fig.~\ref{fig:fig9}) is significantly asymmetric with respect to the centre of this dSph and differs from a spherical or elliptical distribution. The asymmetry of the young stars has already been pointed out (e.g. Stetson et al. 1998; B06), who also find that the older stellar populations are mostly symmetric. The asymmetry is unlikely due to observational biases and may have important implications for the formation mechanisms of these young stars/BSS candidates as well as for their age. In fact, the dynamical crossing time $t_{\rm cross}$ for the BSSs within $r_c$ is $\lesssim{}100$ Myr. The asymmetry should disappear over a time $\gtrsim{}t_{\rm cross}$. On the other hand, the CMD of the Fornax dSph does not show the presence of stars so bright to be associated to such a young ($<$100 Myr old) population; if the BSS candidates here considered are young MS stars most of them must be older than $\gtrsim{}0.5$ Gyr according to their location on the CMD. This result is challenging for the evolution of young stars in Fornax, because a mechanism is needed to keep this asymmetry present on a 
time scale much longer than the crossing time. 

 The left-hand panel of Fig.~\ref{fig:fig9} indicates that the distribution of BSS candidates in Sculptor is more symmetric. However, there is a small asymmetry between the western and the eastern part of the galaxy: the ratio between BSSs in the western part (negative Right Ascension in our plot) and BSSs in the eastern part (positive Right Ascension in our plot) with respect to the centre is ${\rm BSSs(west)}/{\rm BSSs(east)}\sim{}1.4\pm{}0.1$. This asymmetry is also present, although less strong,  in the overall stellar population, since, when we consider all the stars above the 50 per cent completeness limit (i.e. with $V\le{}23.0$ and $I\le{}22.2$), the ratio between stars to the west of the centre and stars to the east of the centre is ${\rm stars(west)}/{\rm stars(east)}\sim{}1.04\pm{}0.01$. 

 The asymmetry in Sculptor is unlikely a spurious effect due to different depths of different WFI pointings. In fact, the asymmetry still persists when we look  at the central pointing alone (i.e. approximately at an area of 34'$\times{}$34' around the centre-of-mass of Sculptor). In the central pointing, we find ${\rm BSSs(west)}/{\rm BSSs(east)}\sim{}1.3\pm{}0.1$ and ${\rm stars(west)}/{\rm stars(east)}\sim{}1.14\pm{}0.02$. 
Thus, we can conclude that this asymmetry between western and eastern side of Sculptor might be real, although its statistical significance is quite low.
Assuming that this asymmetry is real, the mechanisms which may produce it are unclear and deserve further study. For example, is Sculptor tidally perturbed? Alternatively, can a very massive stellar cluster or substructure have been stripped inside Sculptor and have produced the asymmetry? Here, we simply note that there are some observational hints that Sculptor might have swallowed a globular cluster (B07).



\section{Comparison of Sculptor and Fornax with Draco and Ursa Minor}
 Paper~I  showed that BSS candidates in Draco and  Ursa Minor behave like mass-transfer BSSs. In this paper we have seen that the BSS candidates of Sculptor present similarities with respect to BSS candidates in Draco and  Ursa Minor, and that BSS candidates of Fornax are very different from all the other three considered galaxies.
First, the radial distribution of BSSs in Draco and Ursa Minor is similar to that of both RGB and HB stars. The radial distribution of BSS candidates in Sculptor is also similar to that of RHB and RGB stars. Instead, the radial distribution of BSS candidates in Fornax is significantly  more concentrated than that of the overall stellar population (RGB) and those of HB, RC and BL.
Second, in Draco, Ursa Minor and Sculptor there is no significant correlation between the radial distribution and the luminosity distribution. Instead, in Fornax BSS candidates are brighter in the centre.
 Third, the theoretical model for mass-transfer BSSs, adopted in the simulations, can reproduce the radial distribution of BSS candidates in Draco, Ursa Minor and Sculptor (although with higher $\chi{}^2$ than for Draco and Ursa Minor), but has many problems when applied to Fornax BSS candidates.
Fourth, the spatial distribution of BSS candidates in Draco and Ursa Minor is perfectly symmetric with respect to the centre of the dSph, as one would expect for mass-transfer BSSs, whereas  the spatial distribution of BSS candidates in Sculptor and especially in Fornax shows some asymmetry.

Can this comparison help establishing whether or not BSS candidates in Sculptor and Fornax are real BSSs?  
The large differences between the BSS candidates of Fornax and those of the other considered dSphs, together with the arguments presented in Sections $3-5$, disfavour the hypothesis that all the BSS candidates in Fornax are real BSSs. It may be that most BSS candidates in Fornax are young MS stars and the remaining are real BSSs. It would be interesting to look at the spectroscopic differences between BSS candidates hosted in the metal-poor globular clusters of Fornax, which are likely real BSSs, and BSS candidates hosted in the field of Fornax.
In the case of Sculptor,  the radial distribution of BSS candidates is consistent with a population of `real' BSSs, which are coeval to the RHB and to the red RGB. This fact, together with the results for the luminosity distribution and for the simulations, favours the identification of BSS candidates in Sculptor with `real' BSSs, although it is not a definitive evidence. 
For BSS candidates of both Sculptor and Fornax only a spectroscopic analysis, although challenging, may provide decisive results.

\section{Summary}
The existence of BSSs in dSphs is still an open issue and it may have a crucial impact  on our understanding of the star formation history of these galaxies.
In this paper we looked at the BSS candidates in Sculptor and Fornax, in order to understand whether they are real BSSs or young MS stars. We considered photometric observations and compared them with simulations. 

 In Sculptor the observed radial distribution of BSS candidates is similar to the one of RHB stars and is more concentrated than that of BHB stars. Thus, BSS candidates are likely associated with the red, metal-rich population. The resemblance between BSSs and RHB  also favours the identification of BSS candidates with real mass-transfer BSSs, as already seen in Draco and Ursa Minor. In summary, BSS candidates in Sculptor are consistent with real BSSs formed via mass-transfer, although they show some peculiarities. This result supports the hypothesis that star formation  did not occur in Sculptor in recent epochs.


On the contrary, in Fornax the radial distribution of observed BSS candidates is more concentrated than the distribution of all the other considered populations, including ancient (RHB and BHB), intermediate-age (RC) and young stars (BL). This result can hardly be explained by the mass-transfer scenario of BSS formation. The simulations give an acceptable fit of the observed radial distribution only when  all the BSSs are assumed to be born within the core radius $r_c$. The reason why BSS candidates in Fornax require  these peculiar initial conditions is likely that most of them are young stars. 
Furthermore, the luminosity distribution of observed BSS candidates shows a statistically significant trend: internal BSSs are generally brighter than the external ones. This correlation cannot be explained by mass-transfer BSS models. 
In conclusion, Fornax BSS candidates cannot be explained with the mass-transfer model for BSS formation. 
 More likely, BSS candidates in Fornax (or at least most of them) are young MS stars. This result is consistent with the statement, generally accepted in the literature (Stetson et al. 1998; Buonanno et al. 1999; Saviane et al. 2000; Pont et al. 2004; B06), that Fornax hosts young MS stars. 
Furthermore, this finding indirectly strengthens the hypothesis that BSS candidates in Draco, Ursa Minor (paper~I) and probably Sculptor, which show a completely different behaviour from the BSS candidates in Fornax, are real BSSs.

\section {Acknowledgments}
We thank the referee for the critical reading of the manuscript.
MM and ER thank the Kapteyn Astronomical Institute of the
University of Groningen, and
the Institute for Theoretical Physics of the University of Z\"urich for
the hospitality during the preparation of this paper.
MM acknowledges support from the Swiss
National Science Foundation, project number 200020-117969/1
(Computational Cosmology \&{} Astrophysics).

\appendix


\section{A test of the young star hypothesis through isochrones}

\begin{figure*}
\center{{
\epsfig{figure=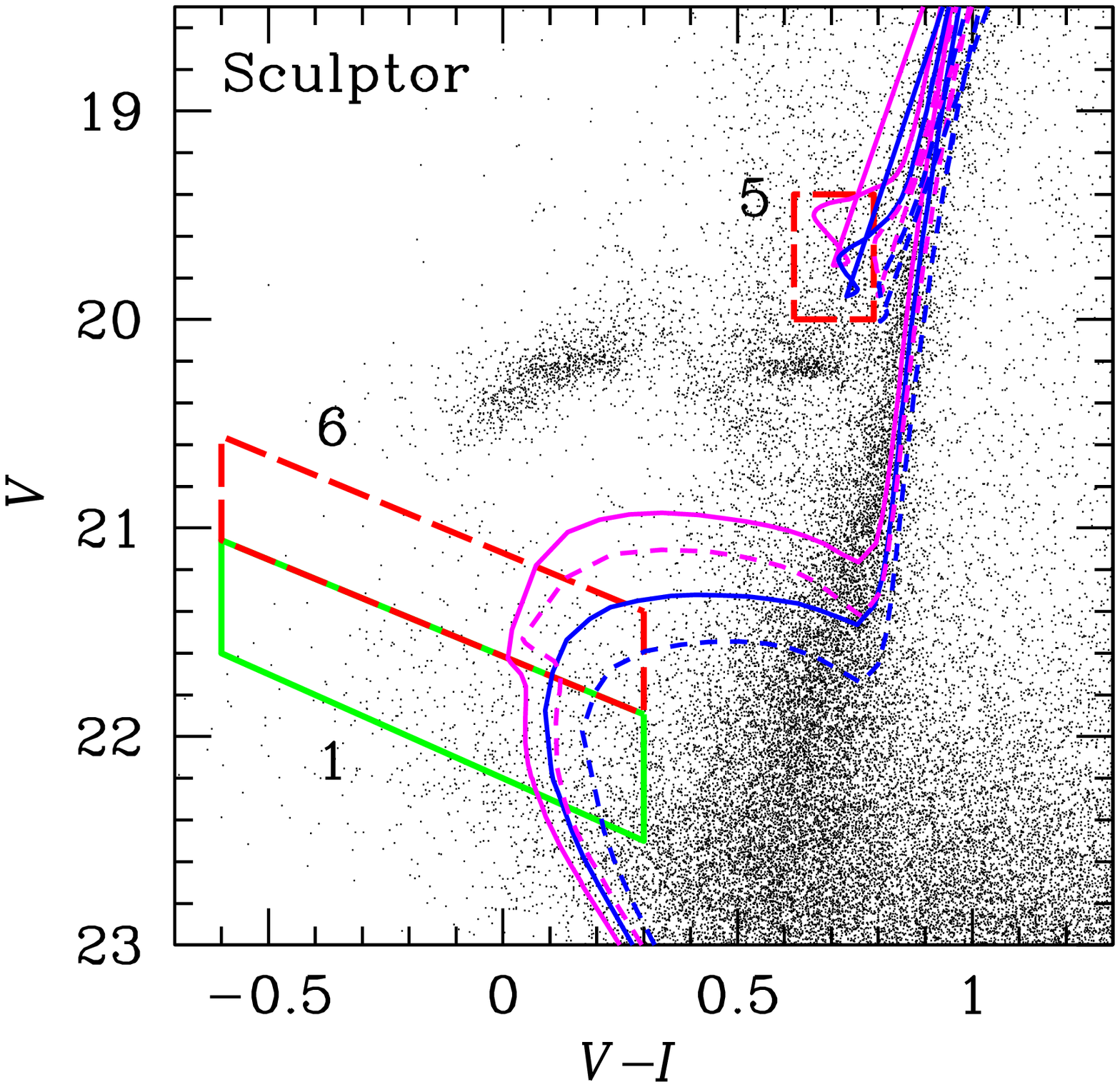,height=8cm}
\epsfig{figure=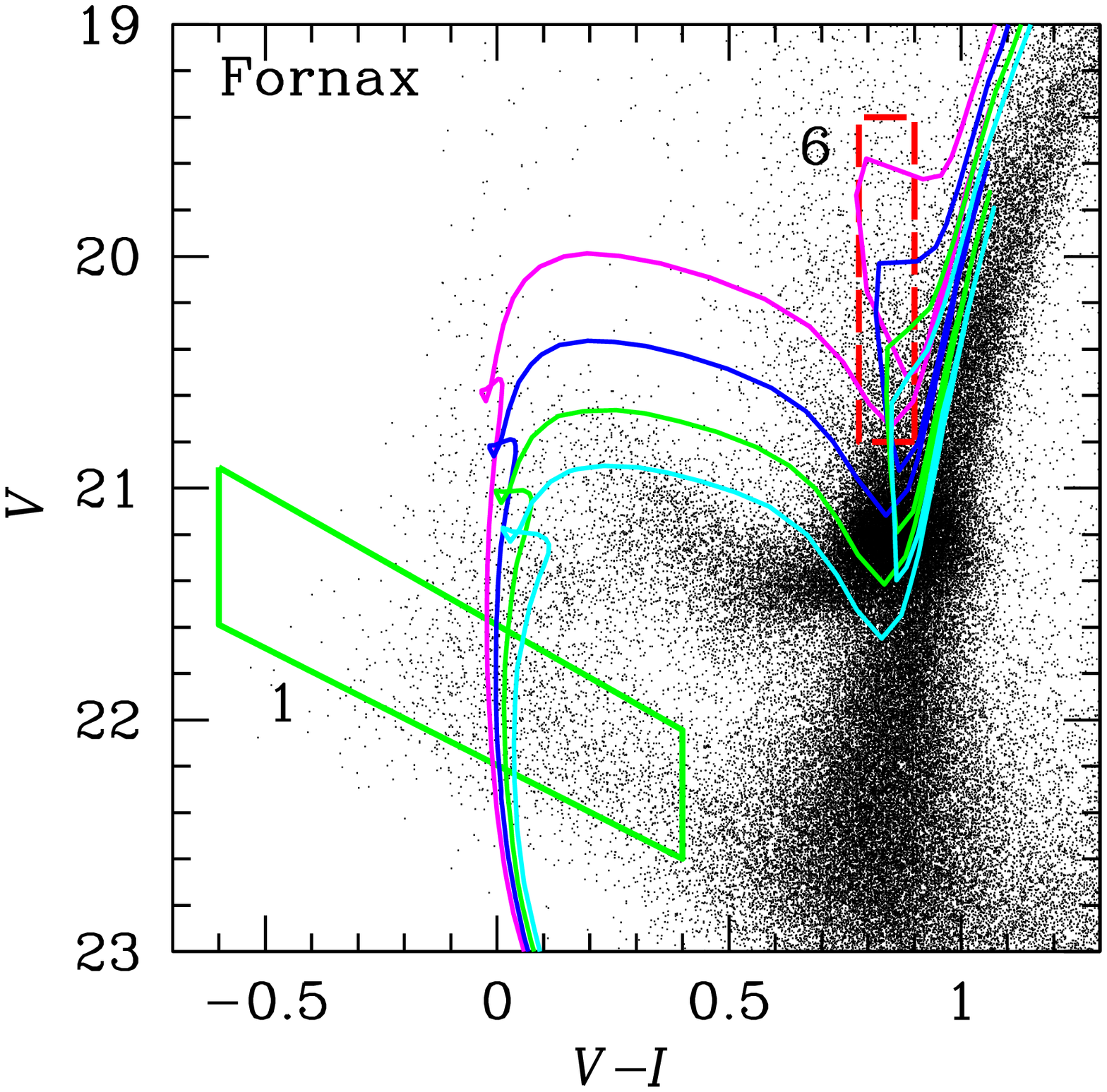,height=8cm}
}}
\caption{\label{fig:figc1} Reddening and distance corrected isochrones
of single stellar populations superimposed to the CMD of the region
within the tidal radius of Sculptor (left-hand panel) and of Fornax (right-hand panel). In the Sculptor plot, solid lines refer to a metallicity
[Fe/H]$=-2.0$, dashed lines to a metallicity [Fe/H]$=-1.5$; the
isochrones refer to ages of 2 Gyr (leftmost pair) and 3 Gyr (rightmost
pair). In the Fornax plot we show isochrones for [Fe/H]$\sim-0.7$
(i.e. Z=0.004); their ages are (from top to bottom) 0.4, 0.5, 0.6, and
0.7 Gyr.
The panels also show the selection boxes for BSSs (solid; labeled 1 in both
panels) and the selection boxes where number counts were made (dashed;
labeled 5 and 6 in the Sculptor CMD, and 6 in the Fornax CMD).}
\end{figure*}


As we already did in paper I, we tested the hypothesis that BSS candidates 
are young stars by checking stellar number counts in
sensitive parts of the CMD, using the isochrones of the Padova group
(see Marigo et al. 2008; see also Girardi et al. 2002, and {\tt
http://stev.oapd.inaf.it/cgi-bin/cmd}), combined with a Chabrier (2003)
log-normal initial mass function.


\subsection{Sculptor}

When theoretical isochrones are plotted over the Sculptor CMD (corrected
for distance and reddening effects; see Fig.~\ref{fig:figc1}), it is
clear that the
BSSs might be associated with a population with an age of 2-3
Gyr, and an average metallicity $-2.0\leq $[Fe/H]$\leq-1.5$ (compatible
with measurements by B07 - fig. 4.8).

More detailed comparisons of number counts provide some hints against
such interpretation; but the evidence is weak, and this method cannot
rule out the possibility that BSSs are actually intermediate-age stars.


The isochrones can also be used to give an indicative estimate
of the upper/lower limit mass of BSSs, which are used to set up our
simulations (see Section 5). For Sculptor, we find that their masses should
be in the range $1.09-1.33\,{}\Msun$.

Finally, we note that if all the BSSs are actually part of a population
with an age of $\sim 2-3$ Gyr, the total mass of such population would
be $\sim 5\times10^4\,{}M_\odot$; if such a star formation episode lasted for
$\sim 1$ Gyr, the implied star formation rate is $\sim 5\times10^{-5}
\,{}M_\odot$ yr$^{-1}$. Such values are larger than what can be inferred in
both Draco and Ursa Minor by a factor of $\sim 5$ (see paper I).

\subsection{Fornax}

The CMD of Fornax is far more complex than the one of Sculptor: the
right-hand panel of Figs.~\ref{fig:fig2} and \ref{fig:figc1}
show a superposition
of old ($\sim 10$ Gyr), intermediate ($2-8$ Gyr) and young ($\lesssim{}1$
Gyr) stellar populations (see B06 and references therein).

The task of disentangling the various
populations appears prohibitive. However, we can at least investigate
whether the number of BSSs is compatible with the number and
distribution of BL stars, in the hypothesis that both are due to the
same young population; the metallicity estimate from B06
($Z\simeq0.004\simeq0.2\,{}Z_\odot$) is very useful in this respect.
Here, we consider isochrones with such metallicity, and ages of 0.4, 0.5,
0.6, and 0.7 Myr.

If we look at the total number of observed BL stars, we find that it is
at least a factor of 2 larger than what can be predicted combining the
observed number of BSSs with any considered isochrone.  In the fainter
half of the BL selection box this might be explained by contamination
from RGB; but similar problems remain also in the brighter half
of the BL selection box.



On the other hand, if we look at the distribution of observed stars within the BL
selection box, `young' (0.4 and 0.5 Gyr) isochrones provide a better fit
than `old' (0.6 and 0.7 Gyr) isochrones. In particular, a 0.4 Gyr
population is required in order to explain the number of stars at the
top of the BL box.


As was done for Sculptor, the isochrones can be used to give an indicative
estimate of the upper/lower limit mass of BSS candidates, and to estimate 
the total mass of the
young population. In the case of Fornax, we find that the masses of stars
in the BSS region (when interpreted as young stars) should be in the
range $1.8-2.3\,{}\Msun$. Furthermore, in order
to explain all the observed BSS candidates, the postulated young population 
should have a total mass of $2-3 \times10^5\,{}M_\odot$, corresponding to a star
formation rate of $\sim{}10^{-3}\,{}\Msun$ yr$^{-1}$ (if the age spread is $\sim
2\times 10^8$ yr): both values are much larger than what we find in
Sculptor, Draco and Ursa Minor.

We conclude that the isochrone method has some difficulties in
explaining all the BSS candidates as part of a young population.
Such difficulties are small when compared to those encountered by
the opposite hypothesis that all BSS candidates are actually BSSs, and
in Fornax the young star hypothesis should be preferred. However, there
remains some space for a small fraction of BSS candidates to be actual
BSSs.

\end{document}